\documentclass[prd,preprintnumbers,floatfix,aps,nofootinbib,notitlepage,showpacs,amssymb,eqsecnum,twocolumn]{revtex4-1}
\usepackage{amsmath}
\usepackage{graphicx}
\usepackage{epstopdf}
\usepackage{latexsym}
\usepackage{amssymb}
\usepackage{color}
\usepackage{bm}
\usepackage{amssymb}
\usepackage{enumitem}
\usepackage{hyperref}
\usepackage{subfigure}
\usepackage{array}
\usepackage[english]{babel}
\usepackage{tensor}
\allowdisplaybreaks

\allowdisplaybreaks

\def\be{\begin{equation}}
\def\ee{\end{equation}}
\def\ba{\begin{eqnarray}}
\def\ea{\end{eqnarray}}
\def\l{\left}
\def\r{\right}
\def\f{\frac}

\def\nn{\nonumber}

\begin{document}

\title{New scaling solutions in cubic Horndeski theories}
\author{In\^es S.~Albuquerque$^1$, Noemi Frusciante$^1$, Nelson J.~Nunes$^1$, and Shinji Tsujikawa$^2$ }
\smallskip
\affiliation{$^1$Instituto de Astrof\'isica e Ci\^encias do Espa\c{c}o, 
Faculdade de Ci\^encias da Universidade de Lisboa,  Campo Grande, PT1749-016 Lisboa, Portugal\\
\smallskip
$^2$Department of Physics, Faculty of Science, Tokyo University of Science,
1-3, Kagurazaka, Shinjuku-ku, Tokyo 162-8601, Japan}

\begin{abstract}

We propose a viable dark energy scenario in the presence of 
cubic Horndeski interactions and a standard scalar-field 
kinetic term with  two exponential potentials. 
We show the existence of new scaling solutions along 
which the cubic coupling $G_3$ provides an important contribution to the field density that scales in the 
same way as the background fluid density.
The solutions finally exit to the epoch of cosmic acceleration 
driven by a scalar-field dominated fixed point 
arising from the second exponential potential.
We clarify the viable parameter space in which 
all the theoretically consistent conditions 
including those for the absence of ghost and Laplacian 
instabilities are satisfied on scaling and scalar-field 
dominated critical points.
In comparison to Quintessence with the same scalar potential, we find that the cubic coupling gives rise to 
some novel features:
(i) the allowed model parameter space is wider 
in that a steeper potential can drive the cosmic acceleration;
(ii) the dark energy equation of state $w_{\phi}$ today 
can be closer to $-1$ relative to Quintessence;
(iii) even if the density associated with the cubic coupling 
dominates over the standard field density
in the scaling era, the former contribution 
tends to be suppressed at low redshifts. 
We also compute quantities associated with the growth 
of matter perturbations and weak lensing potentials 
under the quasi-static approximation in the sub-horizon limit 
and show that the cubic coupling leads to the modified 
evolution of perturbations which can be distinguished 
from Quintessence.

\end{abstract}


\maketitle

\section{Introduction}

The late-time cosmic acceleration has been confirmed by many cosmological surveys, but a satisfactory theoretical explanation for this phenomenon is still lacking. 
Despite the overall success of the 
$\Lambda$-cold-dark-matter model 
in fitting the cosmological data~\cite{Ade:2015xua,Abbott:2017wau}, there are still some shortcomings such as the  cosmological constant and coincidence 
problems \cite{Weinberg:1988cp,Carroll:2000fy,Weinberg:2000yb,Padilla:2015aaa}.  An alternative explanation 
to the cosmic acceleration is to introduce extra fields which modify the gravitational interaction at large 
distances \cite{CST,SF08,DT10,Sotiriou:2010wn,Tsujikawa:2010zza,Clifton:2011jh,Joyce:2014kja,Koyama:2015vza,Heisenberg18}.

Inclusion of a scalar field $\phi$ (or multiple scalar fields)
in the description of the cosmological dynamics 
sometimes gives rise to so-called \textsl{scaling solutions} \cite{Wetterich:1987fm,Peebles:1987ek,Ratra:1987rm,Chiba97,Copeland:1997et,Ferreira:1997hj,Liddle:1998xm,Zlatev:1998tr,
Barreiro:1999zs,Guo:2003rs,Guo:2003eu,Piazza:2004df,Tsujikawa:2004dp,Pettorino:2005pv,Amendola06,Ohashi09,Gomes:2013ema,Chiba:2014sda}. 
The scaling solution is featured by a constant ratio between the energy density of matter components and that of 
the scalar field, in which case there is a possibility for alleviating the coincidence problem. 
Since the field density is not negligibly 
small compared to the background density even in the 
early cosmological epoch, the model can be compatible 
with the energy scale associated with particle physics. 
Moreover, the scaling solution attracts background 
trajectories with different initial conditions. 
After the solutions enter the scaling regime, the cosmological dynamics is completely fixed by theoretical parameters.

In Quintessence described by the Lagrangian 
$G_2=X-V(\phi)$, 
where $X=-\partial_{\mu} \phi \partial^{\mu} \phi/2$ is the standard kinetic term and $V(\phi)$ is the scalar potential, 
there exists a scaling solution for the exponential 
potential $V(\phi)=V_0 e^{-\lambda \phi}$ ($V_0$ and 
$\lambda$ are constants) \cite{Wetterich:1987fm,Copeland:1997et,Ferreira:1997hj,Chiba97,Barreiro:1999zs,Guo:2003rs,Guo:2003eu,Amendola:2014kwa}.
In K-essence given by the Lagrangian $G_2=G_2(\phi, X)$, 
it was shown in Refs.~\cite{Piazza:2004df,Tsujikawa:2004dp} that the condition for the existence of scaling solutions restricts 
the Lagrangian to the form $G_2=Xg(Y)$, where $g$ is an 
arbitrary function of $Y=X e^{\lambda \phi}$.
For example, this includes the diatonic ghost condensate 
model $G_2=-X+ce^{\lambda \phi}X^2$ proposed in 
Ref.~\cite{Piazza:2004df} ($c$ is a constant), 
which corresponds to the choice $g(Y)=-1+cY$.

A further generalization of the Lagrangian including a cubic term $G_3(\phi,X)\square \phi$ can still give rise to 
scaling solutions for the function $G_3=a_1Y + a_2Y^2$ 
(with $a_1,a_2$ constants), along with an exponential 
potential \cite{Gomes:2013ema} and a direct coupling between matter and scalar fields. Finally, another option of having scaling solutions is to consider a conformal coupling to the Ricci scalar. In Ref.~\cite{Pettorino:2005pv}, the authors classified the possible couplings  and selected three forms:  exponential, polynomial, and exponential of polynomial functions. All of them depend on the coefficient 
characterizing the scaling. 

The aim of this work is to explore the possibility for obtaining 
new scaling solutions from cubic Horndeski 
theories \cite{Horndeski:1974wa,Horn1,Horn2,Horn3}. 
For this purpose, we employ the action  
\be
\label{Galileonaction}
{\cal S}= \int d^{4}x \sqrt{-g}\l[\f{R}{16\pi G_N}
+G_2(\phi,X)-G_{3}(\phi,X)\square \phi\r], 
\ee
where $g$ is the determinant of metric tensor $g_{\mu\nu}$, $R$ is the Ricci scalar, and $G_N$ is the Newton gravitational constant. This model is usually denoted as a kinetic gravity braiding model~\cite{Deffayet:2010qz}. The name follows from the presence of a braiding term characterizing the 
mixing of kinetic terms between the scalar field and metric. 
In particular, it arises from the dependence of $G_3$ with 
respect to $X$, i.e., $G_{3,X} \equiv \partial G_{3}/\partial X\neq 0$. 

The phenomenology exhibited by the scalar field through the braiding makes the model attractive as a dark energy candidate.  Indeed, the speed of propagation for the scalar mode gets modified as well as the kinetic term of scalar 
perturbations \cite{Deffayet:2010qz}. 
The braiding term  affects the growth of perturbations, modifies the shape of the matter power spectrum, and the low-$\ell$ tail in the observed Cosmic Microwave Background (CMB) spectrum, thus showing detectable signatures~\cite{Kimura:2010di,Kimura:2011td,Bellini:2014fua}. In cubic Hornseski theories given by the action 
(\ref{Galileonaction}), the propagation speed $c_t$ of 
tensor perturbations on the cosmological background 
is equivalent to that of light \cite{Horn2,DeFelice:2011bh,Lom,Creminelli:2017sry,Ezquiaga:2017ekz,Baker:2017hug,Sakstein:2017xjx, Bettoni:2016mij}. 
Hence  the theories are consistent with the observational 
bound of $c_t$ constrained from the 
gravitational-wave event GW170817 and its electromagnetic 
counterpart \cite{TheLIGOScientific:2017qsa,Monitor:2017mdv}. 

We note that, for some specific choices of cubic Horndeski interactions like covariant Galileons \cite{Coga,DTga}, 
their dominance over other energy densities in the late Universe is not favored by the observational data of galaxy and 
Integrated-Sachs-Wolfe (ISW) 
correlations \cite{Kimura:2011td,Renk}. 
However, it is expected that the observational 
constraints arising from the galaxy-ISW correlation 
generally depend on the form of cubic 
Horndeski interactions.
In particular, unlike covariant Galileons, 
there should exist models in which the density 
associated with the cubic Horndeski term is
subdominant to that of the standard field density.

In this paper,  we construct a dark energy model 
in which the density of cubic Horndeski coupling 
gives important contributions to the field density in scaling radiation and matter eras, but it starts to be subdominant relative to the density arising from 
the standard field Lagrangian $G_2=X-V(\phi)$.
For this purpose, we take into account the sum of two exponential potentials 
$V(\phi)=V_1e^{-\beta_1 \phi}+V_2e^{-\beta_2 \phi}$ 
with $\beta_1 \gg {\cal O}(1)$ and 
$\beta_2 \lesssim {\cal O}(1)$ \cite{Barreiro:1999zs},
 besides the cubic 
Horndeski term $G_3=A \ln Y$ ($A$ is constant) 
allowing for the scaling behavior.
While the exponential potential $V_1e^{-\beta_1 \phi}$ contributes to the field density 
in the early epoch together with the cubic Horndeski 
term, the potential $V_2e^{-\beta_2 \phi}$ dominates over other densities at late time.
We leave the detailed analysis about the compatibility of this model with observational data for a future work, but we show that the quantities $\mu$ and $\Sigma$ associated with Newtonian and weak lensing gravitational potentials \cite{Bean:2010zq,Silvestri:2013ne} 
can be consistent with the conjecture made in 
Ref.~\cite{Pogosian,Peirone:2017ywi}.

This paper is organized as follows. 
In Sec.~\ref{Sec:DynSys}, we present a suitable choice 
of dimensionless variables serving to study the dynamical 
system of background equations of motion in cubic Horndeski theories. We also discuss theoretically consistent conditions 
constrained from the background and the stability of perturbations.
In Sec.~\ref{Sec:Aconst}, we choose a specific form of 
the cubic coupling $G_3$ allowing for scaling solutions and study the corresponding critical points and their stability. In Sec.~\ref{Sec:Cosmology}, we show a practical example of realizing scaling radiation/matter eras followed 
by the epoch of cosmic acceleration and study the 
background cosmological dynamics in detail. 
In Sec.~\ref{persec}, we discuss the evolution of quantities 
relevant to Newtonian and weak lensing 
gravitational potentials.
Finally, we conclude in Sec.~\ref{Sec:Conclusions}.

\section{Dynamical system and stability}
\label{Sec:DynSys}

We derive the background equations of motion 
in cubic Horndeski theories on the flat 
Friedmann-Lema\^{i}tre-Robertson-Walker (FLRW) spacetime and construct the dynamical system 
by choosing an exponential potential for the scalar field. 
We also apply conditions for the absence of ghost and Laplacian instabilities derived in 
Ref.~\cite{DeFelice:2011bh} to our cubic Horndeski theories.

The first step of our analysis is to write the corresponding background field equations as an autonomous system of first-order differential equations. Afterwards, the cosmological dynamics is determined  by investigating the evolution around critical points. 
The stability of the critical points is known by linearizing 
the autonomous equations around each point and 
computing eigenvalues of the Jacobian matrix associated to the system. A critical point is stable when all the eigenvalues are negative and the point is said to be a \textsl{stable node} or \textsl{attractor}; it is unstable when the eigenvalues are all positive and the point is said to be an \textsl{unstable node}; finally, when at least one eigenvalue is positive and one negative, the point is a \textsl{saddle}.  We refer the reader to  Refs.~\cite{Wainwright,Strogatz} for details on the procedure and to Refs.~\cite{Copeland:1997et,Baccigalupi:2000je,Matarrese:2004xa,CST,Zhou:2009cy,Leon:2012mt,Mueller:2012kb,Frusciante:2013zop,Paliathanasis:2015gga,vandeBruck:2016jgg,Dutta:2017fjw,Bahamonde:2017ize,SantosdaCosta:2018ovq} for applications to alternative theories of gravity. 

Let us consider the action~(\ref{Galileonaction}) in the 
presence of matter perfect fluids described by 
the action $\mathcal{S}_{\gamma}$. 
To study the background equations of motion, we 
use the unit $8\pi G_N=1$.
On the flat FLRW background given by 
the line element $ds^2=-dt^2+a^2(t) \delta_{ij}dx^i dx^j$, 
where $a(t)$ is a time-dependent scale factor, 
the modified Friedmann equations are
\ba
& &
3H^{2} = \rho_{\phi}+ \rho_{\gamma}\,, \label{Freq} \\
& &2\dot{H} + 3H^{2} =-p_{\phi} -p_\gamma \,, \label{evoeq}
\ea
where a dot represents the derivative with respect to $t$, 
$H=\dot a/a$ is the Hubble expansion rate, 
$X=\dot{\phi}^2/2$, and 
\ba
\rho_{\phi} &=& 2XG_{2,X} - G_2 
+ 6 X \dot \phi H G_{3,X} 
- 2XG_{3,\phi}\,,\\
p_{\phi} &=&  G_2-2X(G_{3,\phi} 
+\ddot{\phi}\,G_{3,X})\,.
\ea
The quantities $\rho_{\phi}$ and $p_{\phi}$ correspond to 
the density and pressure arising from the scalar field, 
respectively, whereas $\rho_{\gamma}$ 
and $p_\gamma$ are those of perfect fluids. 
The fluid equation of state is given by 
$\gamma-1=p_\gamma/\rho_{\gamma}$, where 
$\gamma$ is a constant barotropic coefficient in 
the range $0<\gamma<2$.
The dust corresponds to the choice $\gamma=1$ and radiation to $\gamma=4/3$.

Variation of the action with respect to $\phi$
leads to 
\be
\label{scalareq}
\frac{1}{a^3} \frac{d}{dt} \left( a^{3} J 
\right)=P\,,
\ee
where 
\ba
&&J= \dot{\phi}\,G_{2,X} + 6HX G_{3,X} 
- 2\dot{\phi}\,G_{3,\phi}\,,\nn\\
&&P=G_{2,\phi} -2X \left(G_{3,\phi \phi} 
+ \ddot{\phi}\,G_{3,\phi X} \right) \,.
\label{Pdef}
\ea

In the following, we consider the quadratic 
Lagrangian $G_2(\phi,X)$ of a standard canonical 
scalar field, i.e., 
\be
G_2(\phi,X)=X-V(\phi)\,,
\ee
with an exponential potential 
\be
V(\phi)=V_{0} e^{-\beta \phi}\,,
\label{Vexp}
\ee
where $V_0$ and $\beta$ are constants. 

In the search for scaling solutions, let us now write $G_3$ 
in the form 
\be
G_3(\phi,X)=g(Y)\,,\qquad Y=Xe^{\lambda \phi}\,, 
\ee
where $g$ is an arbitrary function of $Y$, 
and $\lambda$ is a constant.
With this definition, the derivatives of $G_3$ with respect to 
$\phi$ and $X$ can be written as 
\begin{flalign}
&G_{3,\phi} = \lambda Y g_{,Y} \ ,\label{dphi} \\
&G_{3,X} = \frac{Y g_{,Y}}{X}\ , \label{dx} \\
&G_{3,XX} = \frac{Y^2g_{,YY}}{X^2} \ , \label{dxx} \\
&G_{3,\phi \phi} = \left( g_{,Y}+Yg_{,YY} \right)   
\lambda ^2 Y \ , \\
&G_{3,\phi X} = G_{3,X \phi} 
= \left( g_{,Y}+Yg_{,YY} \right)
\lambda \frac{Y}{X}\,.  
\label{G3phiX}
\end{flalign}
To study the background cosmological dynamics, we 
introduce the dimensionless variables:
\be
x = \frac{\dot{\phi}}{\sqrt{6} H}, \qquad y 
= \frac{\sqrt{V}}{\sqrt{3}H} \,, \qquad 
\Omega_{\gamma}= \f{\rho_{\gamma}}{3H^2}\,.
\ee
On using Eqs.~(\ref{Freq})-(\ref{Pdef}) and 
the relations (\ref{dphi})-(\ref{G3phiX}), 
we obtain the following autonomous system of 
first-order differential equations:
\ba
\label{xprime}
x' &=& \frac{1}{\sqrt{6}} f(x,y) - \frac{\dot{H}}{H^2}x, \\
\label{yprime}
y' &=& - \sqrt{\frac{3}{2}} \beta x y - \frac{\dot{H}}{H^2}y\,,
\ea
where a prime represents the derivative with respect to 
$N=\ln a $, and 
\ba
f(x,y) &\equiv&\frac{\ddot{\phi}}{H^2} \nonumber \\
&=& 
s(x,y) \left\{ -3 \sqrt{6} x + 3 \beta y^2 \right. \nonumber \\
&~&-6 \lambda x \left( Yg_{,Y} + Y^2g_{,YY} \right)
(\sqrt{6} - \lambda x)  \nonumber \\ 
&~& -6Yg_{,Y} \left[ 3-3 x^2- \sqrt{6} \lambda x 
- \frac{3}{2} \gamma\,\Omega_{\gamma} \right. \nonumber \\
&~& \left. \left. +3xYg_{,Y}(2\lambda x - \sqrt{6})
\right] \right\},
\label{fxy}
\ea
with
\ba
s(x,y)^{-1} &=& 1+2\left( Yg_{,Y} + Y^2g_{,YY} \right) 
\left(\frac{\sqrt{6}}{x} - \lambda \right)  \nonumber \\
&&+ 2 Y g_{,Y} \left( 3 Yg_{,Y}  - \lambda \right) \,,
\ea
and
\be
\frac{\dot{H}}{H^2}= Yg_{,Y} f -\frac{3}{2} \gamma 
\Omega_{\gamma} +3x Yg_{,Y}  
\left(2 \lambda  x-\sqrt{6} \right)-3 x^2\,.
\label{dotH}
\ee
The critical points $(x_c, y_c)$ of the above dynamical system 
can be derived by setting $x'=0$ and $y'=0$ 
in Eqs.~(\ref{xprime})-(\ref{yprime}).

The constraint equation (\ref{Freq}) can be expressed as
\be
\label{const}
\Omega_{\gamma}=1-\Omega_{\phi} \,,
\ee
where $\Omega_{\phi}$ is 
the field density parameter defined by 
\be
\label{omecons}
\Omega_{\phi} \equiv 
\frac{\rho_{\phi}}{3H^2}
=x^2 + y^2 + 2 x Y g_{,Y} \left( \sqrt{6}- \lambda x 
\right) \,.
\ee
For a positive fluid density ($\Omega_{\gamma} \geq 0$), 
$\Omega_{\phi}$ has an upper bound, $\Omega_{\phi}\leq 1$. 
This condition will be exploited when exploring 
the region in which the critical points exist. 
We also introduce the density parameter associated with 
the cubic coupling $G_3$, as 
\be
\label{omeG3}
\Omega_{G_3}=2 x Y g_{,Y} 
\left( \sqrt{6}- \lambda x 
\right) \,.
\ee
The scalar-field equation of state $w_{\phi}$ and 
the effective equation of state $w_{\rm eff}$ are 
defined, respectively, by 
\ba
w_{\phi} &\equiv& \frac{p_{\phi}}{\rho_{\phi}} 
= \frac{x^2 - y^2 - 2Yg_{,Y}(f/3+ \lambda x^2)}
{x^2 + y^2 + 2 Yg_{,Y} ( \sqrt{6}x-\lambda x^2)}\,,
\label{wphi}\\
w_{\rm eff} &\equiv& -1-\frac{2\dot{H}}{3H^2}\,.
\ea
The Universe exhibits the accelerated expansion for 
$w_{\rm eff}<-1/3$. In general, the condition 
$w_{\phi}<-1/3$ is not sufficient for realizing 
the cosmic acceleration.
If the energy density of the Universe is dominated by 
the scalar field, then the cosmic 
acceleration occurs under the condition 
$w_{\phi}<-1/3$.

So far, we have maintained the form of $G_3=g(Y)$ completely open. In order to close the system, however, 
we have to make a choice for this function. 
If we choose $\beta=\lambda$, as done 
in Refs.~\cite{Piazza:2004df,Tsujikawa:2004dp}, 
the quantity $Y$ can be expressed as 
$Y=(x^2/y^2)V_0$ and hence Eqs.~(\ref{xprime})-(\ref{yprime}) are immediately  closed. In the following, we will not make this assumption and explore another possibility to close the dynamical system. 
An explicit model is presented in Sec.~\ref{Sec:Aconst}.

To guarantee the viability of our model at the background level, 
we impose the following conditions:
\begin{itemize}
\item \textit{Existence condition:} 
The critical points $(x_c, y_c)$ must be real.
\item \textit{Stability of critical points:} 
We need to identify critical points responsible 
for radiation/matter eras and for the late-time cosmic 
acceleration. {}From a cosmological point of view, 
the fixed points during the radiation and matter eras need to be either an unstable node or a 
saddle point. The system finally has to approach an attractor/stable point with the cosmic acceleration.
For the late-time scalar-field dominated solution, 
we demand the condition $w_\phi<-1/3$. 
\item \textit{Phase-space constraint:} 
We impose that the field density parameter is in the range $\Omega_{\phi} \leq 1$.
\end{itemize}

In Horndeski theories, there are two tensor and one scalar degrees of freedom. 
In cubic Horndeski theories, the second-order action of tensor perturbations is the same as that in General Relativity (GR) \cite{Horn2,DeFelice:2011bh}, so there are neither ghost nor Laplacian instabilities in the tensor sector. In particular, the  
speed of gravitational waves is equivalent to that of light.
On the other hand, the no-ghost condition and the sound speed 
$c_s$ of scalar perturbations get modified by cubic Horndeski terms 
compared to a canonical scalar field with the Lagrangian 
$G_2=X-V(\phi)$.
\begin{itemize}
\item \textit{Physical viability conditions:\footnote{We do not use the more appropriate wording ``physical stability conditions'' in order to avoid confusion with the ``stability conditions'' of critical points which determine the dynamics of the system.}}
The conditions for the absence of ghost and 
Laplacian instabilities in the small-scale limit are given, respectively, by \cite{DeFelice:2011bh}
\ba
Q_s &\equiv&  \frac{4 w_3 + 9 w_2^2}{3 w_2^2} > 0 \,,
\label{Qsdef} \\
\label{laplas}
c_s^2 &\equiv& \frac{2 \left( Hw_2  - \dot{w_2} - \gamma \rho_{\gamma} \right) - w_2^2}{w_2^2 Q_s}>0\,,
\label{csdef}
\ea
where 
\ba
w_2 &=& 2H \left( 1 - Yg_{,Y}  \sqrt{6} x \right) \,,\nn\\
w_3 &= & 9 H^2 \left[x^2 \left( 1-2 \lambda Y^2 g_{,YY} 
- 4 \lambda Yg_{,Y}  \right) \right.  \nonumber \\
& & \left. + 2\sqrt{6}x \left( 2 Yg_{,Y} + Y^2 g_{,YY} \right) 
-1 \right]\,.
\ea
\end{itemize}
For a given model, the viable parameter space is constrained 
to satisfy all the conditions mentioned above.

\section{Model with scaling solutions}
\label{Sec:Aconst}

\begin{table*}[th!]
\small
\begin{center}
\begin{tabular}{|c|c|c|c|c|c|c|}
\hline
 & $x_c$ & $y_c^2$ & $ \Omega_{ \phi }$ & $w_{\phi}$ & $Q_s$& $c_s^2$\\ 
\hline\hline
(a) & $\sqrt{\frac{3}{2}} \frac{\gamma }{\beta }$ & $\frac{3(2-\gamma)}{2\beta^2} [\gamma + 2 A (\beta-\gamma\lambda)] $ & Eq.~(\ref{Omea}) & $ \gamma - 1 $ & $\frac{9\gamma^2}{2} \frac{1+2A(3A-\lambda)}{(\beta-3A\gamma)^2}$ &   $
1+ \frac{8A}{3\gamma} \frac{\beta-3A\gamma}
{1+2A(3A-\lambda)}$    \\ \hline 
(b) & $\frac{\beta-6 A }{\sqrt{6}[1+A (\beta-2\lambda)]}$ & $\frac{[1+2A(3A-\lambda)] [6-\beta ^2+12 A (\beta -\lambda )]}
{6[1+A (\beta -2 \lambda )]^2}$ & 1 & $-1+\frac{\beta (\beta -6 A)}{3[1+A(\beta-2\lambda)]}$ & $ \frac{(\beta -6 A)^2}
{2[1+2A(3A-\lambda)]}$ &  $1+\frac{8 A}{\beta -6 A}$ 
\\ \hline
(c) & $\frac{\sqrt{6} A}{2 A \lambda -1}$ & 0 & $\frac{6 A^2}{2 A \lambda -1}$ &  $ \gamma - 1 $ & $\frac{18 A^2}{1+2A(3A-\lambda)}$ 
&$-\frac{1}{3}$  \\ \hline
(d1) & $\frac{\sqrt{6} A-\sqrt{1+2A(3A-\lambda)}}{2 A \lambda -1}$ & 0 & 1 & 1  &3 &  $1+\frac{4\sqrt{6}A}{3\sqrt{1+2A(3A-\lambda)}}$ \\ \hline
(d2) & $\frac{\sqrt{6} A + \sqrt{1+2A(3A-\lambda)}}{2 A \lambda -1}$ & 0 & 1 & 1 & 3 &  $1-\frac{4\sqrt{6}A}{3\sqrt{1+2A(3A-\lambda)}}$  \\ \hline
\end{tabular}
\end{center}
\caption{\label{critp}
Critical points $(x_c,y_c^2)$ of the dynamical system (\ref{xprime})-(\ref{yprime}) 
for the model given by the functions $G_2=X-V_0e^{-\beta \phi}$ and 
$G_3=A \ln Y$ with $Y=Xe^{\lambda \phi}$, in the presence 
of a barotropic perfect fluid with the equation of state $\gamma-1$.
For each critical point, we also show the values of 
$\Omega_\phi$, $w_\phi$, $Q_s$, and $c_s^2$ defined, 
respectively, by Eqs.~(\ref{omecons}), (\ref{wphi}), (\ref{Qsdef}), and (\ref{csdef}).
The fluid density parameter is known by the relation 
$\Omega_{\gamma}=1-\Omega_{\phi}$.}
\end{table*}

{}From Eqs.~(\ref{xprime})-(\ref{yprime}) 
with Eqs.~(\ref{fxy})-(\ref{dotH}), 
the dynamical system is closed for  
$g(Y)$ satisfying the conditions that both 
$Yg_{,Y}$ and $Y^2g_{,YY}$ are constants.
Let us consider the cubic coupling given by 
\be
\label{G3}
G_3(\phi,X) = g(Y) = A \ln Y\,,
\ee
where $A$ is a constant. 
In this case, we have
\be
Yg_{,Y} = - Y^2 g_{,YY}= A = \mbox{constant}. 
\label{gY}
\ee
In the following, we derive the critical points and discuss the stability 
of them for the cubic coupling (\ref{G3}).

In Table \ref{critp}, we show the critical points and their corresponding 
values of $\Omega_{\phi}$ and $w_{\phi}$. 
The fluid density parameter is known from the relation 
$\Omega_{\gamma}=1-\Omega_{\phi}$. 
For the model (\ref{G3}), the physical viability conditions 
(\ref{Qsdef})-(\ref{csdef}) reduce, respectively, to 
\ba
&&Q_s=\frac{3 x^2 \left(1-2 A \lambda+6 A^2 \right)}
{\left(1-\sqrt{6} A x\right)^2} >0\,,\label{stabAconst}\\
&&c_s^2=\frac{3x(1-2A\lambda-2A^2)+4 \sqrt{6} A}
{3x \left( 1-2 A \lambda+6 A^2\right)}>0\,.
\label{stabAconst2}
\ea
In the limit $A \to 0$, we have  $Q_s \to 3x^2$ and 
$c_s^2 \to 1$, so the conditions $Q_s>0$ and $c_s^2>0$ are 
automatically satisfied. 
In presence of the cubic coupling (\ref{G3}), 
the parameters $A$ and $\lambda$ are constrained to satisfy the conditions (\ref{stabAconst}) and (\ref{stabAconst2}). 
In Table \ref{critp}, we present concrete values of 
$Q_s$ and $c_s^2$ for each fixed point.

In what follows, we will illustrate the main characteristics of the 
critical points and discuss their stability following the criteria 
mentioned in Sec.~\ref{Sec:DynSys}. 
In Appendix \ref{Eigenvalues}, we explain the detail for 
the stability of the fixed points by explicitly computing 
eigenvalues of the Jacobian matrix associated with homogeneous perturbations around each point. 
In total, there are five fixed points presented below and 
in Table~\ref{critp}.
\begin{itemize}
\item Point (a): This point is characterized by 
\be
x_c=\sqrt{\frac{3}{2}} \frac{\gamma }{\beta}\,,\qquad 
y_c=\sqrt{\frac{3(2-\gamma)}{2\beta^2} 
[\gamma + 2 A (\beta-\gamma\lambda)]}\,,
\ee
with $w_{\phi}=w_{\rm eff}=\gamma-1$ and
\be
\Omega_{\phi}=
\frac{3}{\beta^2} \left[\gamma +A \left\{ 
\gamma(\beta-2\lambda)+2\beta \right\} \right]\,.
\label{Omea}
\ee
This corresponds to the {\it scaling solution} along which 
the ratio of energy densities between $\phi$ and the matter fluid 
are constant ($\Omega_\phi/\Omega_\gamma=$ constant) 
with $w_\phi$ equivalent to the matter equation of state ($w_{\phi}=\gamma-1$). 
The dark energy density scales as the fluid density 
regardless of the value of $\gamma$. 
The scaling ratio $\Omega_\phi/\Omega_\gamma$ depends on 
the parameters $\beta, \gamma, A, \lambda$. 
In the limit $A \to 0$, the values of $\Omega_{\phi}$ and 
$y_c$ given above recover those derived for a canonical scalar field with the  exponential potential 
(\ref{Vexp}) \cite{Copeland:1997et}. 
Existence of the cubic coupling (\ref{G3}) modifies 
the ratio $\Omega_\phi/\Omega_\gamma$. 
The density parameter (\ref{omeG3}) arising from the cubic coupling reads
\be
\Omega_{G_3}=\frac{3}{\beta^2}
A \gamma \left( 2\beta-\gamma \lambda \right)\,,
\ee
which gives an important contribution to the field density 
(\ref{Omea}).

For the existence of point (a), we require that $y_c$ is real and hence 
\be
\gamma + 2 A (\beta-\gamma\lambda) \geq 0\,.
\label{acon1}
\ee
{}From the values of $Q_s$ and $c_s^2$ shown in Table \ref{critp}, the ghost and Laplacian instabilities are 
absent under the conditions: 
\ba
& &
1+2A \left( 3A-\lambda \right)>0\,. \label{acon2}\\
& &
8A \beta+3 \gamma \left[ 1-2A(A+\lambda) 
\right]>0\,.
\label{acon3}
\ea
The eigenvalues of the Jacobian matrix for point (a) 
are given by  Eq.~(\ref{aeigen}) in Appendix \ref{Eigenvalues}.
On using the requirements (\ref{acon1}) and (\ref{acon2}) 
as well as the conditions 
$0<\gamma<2$ and $\Omega_{\phi} \leq 1$, 
it follows that neither $\mu_1$ nor $\mu_2$ can be positive. 
This means that the scaling fixed point (a) is always stable 
under theoretically consistent conditions. 
In other words, if one wants to use point (a) to realize the 
scaling solution during the radiation and matter eras, 
one needs to consider an additional mechanism of exiting from 
the scaling regime  to the epoch of cosmic acceleration. 
In Sec.~\ref{Sec:Cosmology}, we will propose 
a concrete model allowing such a possibility.

If point (a) is responsible for the scaling radiation era,  
there is an extra constraint arising from the Big Bang Nucleosynthesis 
(BBN) \cite{Ferreira:1997hj}. 
The field density parameter in the scaling radiation 
era  $\Omega_{\phi}^{(r)} \equiv \Omega_{\phi} (\gamma=4/3)$ is 
constrained to be \cite{Bean}   
\be
\label{earlyDEconst}
\Omega_{\phi}^{(r)}=
\frac{4}{\beta^2} \left[ 
1+\frac{A}{2} \left( 5\beta-4\lambda \right) \right]
<0.045\,.
\ee
The cubic coupling $G_3$ leads to the modification to
the standard value $\Omega_{\phi}^{(r)}=4/\beta^2$
derived for a canonical scalar field with the exponential potential $V(\phi)=V_0e^{-\beta \phi}$. 
Whether $\Omega_{\phi}^{(r)}$ is increased or decreased 
by $G_3$ depends on the sign of $A(5\beta-4\lambda)$.
In 2015, the Planck team \cite{Ade:2015rim} put 
a more stringent bound on the field density parameter 
from CMB measurements: $\Omega_{\phi}<0.02$ (95\% C.L.) 
at the redshift $z \equiv 1/a-1 \approx 50$.
If the solution is in the scaling regime 
during the matter era, the field density parameter 
$\Omega_{\phi}^{(m)} \equiv \Omega_{\phi} (\gamma=1)$ 
is constrained to be 
\be
\label{earlyDEconst2}
\Omega_{\phi}^{(m)}=\frac{3}{\beta^2} \left[ 
1+A\left( 3\beta-2\lambda \right) \right]<0.02\,.
\ee
In Sec.~\ref{Sec:Cosmology}, we will present 
a model with the early-time scaling solution 
followed by the late-time cosmic acceleration. 
We show that it is possible to find the parameter space consistent with all the bounds derived above.

\item Point (b): This point corresponds to 
\ba
\hspace{-0.4cm}
x_c &=&\frac{\beta-6 A }{\sqrt{6}[1+A (\beta-2\lambda)]}\,,
\label{xb}\\
\hspace{-0.4cm}
y_c &=& \sqrt{\frac{[1+2A(3A-\lambda)] 
[6-\beta ^2+12 A (\beta -\lambda )]}{6[1+A (\beta -2 \lambda )]^2}},
\ea
with $\Omega_{\phi}=1$, and 
\be
w_{\phi}=w_{\rm eff}=-1+\frac{\beta (\beta -6 A)}{3[1+A(\beta-2\lambda)]}\,.
\label{wphib}
\ee
This is the scalar-field dominated point which can be used 
for the late-time dark energy.
In this case, the cosmic acceleration occurs for 
$w_{\rm eff}<-1/3$, i.e., 
\be
\frac{\beta (\beta -6 A)}{1+A(\beta-2\lambda)}<2\,.
\label{bcon1}
\ee
{}From the values of $Q_s$ and $c_s^2$ shown in 
Table \ref{critp}, there are neither ghost nor 
Laplacian instabilities under the conditions:
\ba
& &
1+2A \left( 3A-\lambda \right)>0\,,\label{bcon2}\\
& &
\left( \beta+2A \right) \left( \beta-6A \right)>0\,,
\label{bcon3}
\ea
where Eq.~(\ref{bcon2}) is the same as Eq.~(\ref{acon2}).
Since $y_c$ must be real, we require that 
\be
6-\beta ^2+12 A \left( \beta -\lambda \right)
\geq 0\,,
\label{bcon4}
\ee
where we used the condition (\ref{bcon2}).
If we demand that point (b) is the late-time attractor, 
the two eigenvalues (\ref{beigen1}) and (\ref{beigen2}) 
given in Appendix \ref{Eigenvalues} need to be negative, 
so that 
\ba
& &
\mu_1=\frac{\beta(\beta-6A)}
{1+A(\beta-2\lambda)}-3<0\,,
\label{bcon5}\\
& &
\mu_2=
\frac{\beta^2-6+12A(\lambda-\beta)}
{2[1+A(\beta-2\lambda)]}<0\,,
\label{bcon6}
\ea
where we have chosen the value $\gamma=1$ in Eq.~(\ref{bcon5}).

In the limit $A \to 0$, the conditions (\ref{bcon2}) 
and (\ref{bcon3}) are automatically satisfied, while the other 
conditions (\ref{bcon1}), (\ref{bcon4}), (\ref{bcon5}), 
and (\ref{bcon6}) are satisfied for $\beta^2<2$.
This upper bound of $\beta$ is modified by the 
nonvanishing coupling $A$. {}From Eq.~(\ref{wphib}), 
we observe that it is possible to realize 
$w_{\phi} \simeq -1$ for the coupling $A$ close to $\beta/6$. 
We need to caution that the bound (\ref{bcon3}), which 
arises from the condition $c_s^2>0$, places the upper limit 
on the amplitude of $A$. For $\beta>0$, 
this bound translates to
\be
-\frac{\beta}{2}<A<\frac{\beta}{6}\,,
\ee
and hence $A$ cannot be larger than $\beta/6$. 
For point (b), the density parameter (\ref{omeG3}) 
associated with the cubic coupling is given by 
\be
\Omega_{G_3}=
\frac{A(\beta-6A)[6+6A (\beta-\lambda)-\beta \lambda]}
{3[1+A(\beta-2\lambda)]^2}\,,
\label{OmeG3b}
\ee
which vanishes for $A=\beta/6$. 
For $A$ close to $\beta/6$, the cubic coupling
slowdowns the evolution of $\phi$, so that 
$x_c \approx 0$ in Eq.~(\ref{xb}). 
In this case, the dominant contribution to $\Omega_{\phi}$ 
comes from the potential energy, i.e., 
$\Omega_{\phi} \approx y_c^2=1$.

For the model presented in Sec.~\ref{Sec:Cosmology}, 
the late-time cosmic acceleration is driven by point (b).
There exists the viable parameter space 
in which all the conditions 
(\ref{bcon1})-(\ref{bcon6}) are consistently satisfied.

\item Point (c): This is a {\it kinetic scaling solution} which 
exists for $A\neq 0$. Since $y_c=0$, the field potential 
does not play any role. The nonvanishing 
field kinetic energy $x_c=\sqrt{6}A/(2A\lambda-1)$
leads to the constant density parameter 
$\Omega_{\phi}=6A^2/(2A\lambda-1)$. 
Since the scalar propagation speed squared is negative 
($c_s^2=-1/3$), the physical viability condition 
(\ref{csdef}) is not satisfied for $A \neq 0$.
In the limit that $A \to 0$, this fixed point corresponds to a 
fluid-dominated solution ($\Omega_{\gamma}=1$) 
with $c_s^2=1$, see Eq.~(\ref{stabAconst2}). 
For $A=0$, the eigenvalues (\ref{ceigen1})-(\ref{ceigen2}) 
given in Appendix \ref{Eigenvalues}
are in the ranges $\mu_1<0$ and $\mu_2>0$, so 
the fluid-dominated solution corresponds to a saddle point 
which can be used for the radiation or matter era.

We stress that, for $A \neq 0$, point (c) is excluded by the 
negative $c_s^2$, so it can not play the role of scaling 
radiation or matter eras.

\item Points (d1) and (d2): 
These fixed points are {\it kinetically dominated} scalar field solutions where the kinetic energy 
of $\phi$ is the dominant component to the total 
energy density.
One of the eigenvalues $\mu_1=3(2-\gamma)$ is positive, so they are either unstable 
or saddle points. However, since $w_{\rm eff}=w_{\phi}=1$ 
and $\Omega_{\phi}=1$ on these points, they are responsible 
for neither radiation nor matter eras. 
Moreover, it cannot be used for the late-time cosmic acceleration.
\end{itemize}

In summary, we showed that neither points (c) nor 
(d1,d2) are suitable to describe a viable cosmic expansion history after the onset of the radiation-dominated epoch.
The point (a) can be responsible for scaling radiation and
matter eras, but the solution does not exit from the scaling 
regime to the epoch of cosmic acceleration. 
This comes from the property that the scaling solution (a) 
is stable for $\Omega_{\phi}<1$  with the other theoretical consistent conditions (\ref{acon1}) and (\ref{acon2}) . 
{}From Eqs.~(\ref{earlyDEconst}) and (\ref{earlyDEconst2}), we generally require that the value of $|\beta|$ be larger than 
order unity. In this case, the exponential potential is so 
steep that it is also difficult to satisfy all the conditions 
required for the existence and stability of point (b) 
with the cosmic acceleration.
This situation changes for the scalar potential in which 
the slope $\tilde{\beta}=-V_{,\phi}/V$ decreases in time.
In Sec.~\ref{Sec:Cosmology}, we will study a modified version of the present model by including a second potential term. 
As discussed in  Ref.~\cite{Barreiro:1999zs} for standard 
Quintessence, this allows the possibility for realizing scaling 
radiation/mater eras followed by the late-time accelerated expansion.

In Ref.~\cite{Gomes:2013ema}, the authors showed that there exist scaling 
solutions for the cubic coupling $g_3(Y)=c_1 Y+c_2 Y^2$ 
by considering a field-dependent coupling $Q(\phi)$ between 
the scalar field and nonrelativistic matter. 
In addition to the fact that a specific form of the coupling 
$Q(\phi)=(b_1 \phi+b_2)^{-1}$ was chosen, they made some 
additional assumptions for deriving the solution to $g_3(Y)$. 
Moreover, their results are not directly applicable to the 
case $Q=0$. As we explicitly showed above, scaling 
solutions are present even for the 
function $g_3(Y)=A \ln Y$.
Indeed, as we will show in a separate publication, there are 
scaling solutions even for general arbitrary functions $g_3(Y)$.

While the cubic coupling $g_3(Y)=A \ln Y$ is chosen 
in this paper due to its simplicity, 
it can accommodate most of the important properties of scaling 
solutions. Moreover, this is an explicit and simple example of 
showing the existence of scaling solutions other than 
the coupling $g_3(Y)=c_1 Y+c_2 Y^2$. 
In Sec.~\ref{Sec:Cosmology}, we will consider
a double exponential potential for realizing an exit from 
the scaling matter era. In any scaling solution relevant to 
radiation/matter eras, we need a mechanism of transition from 
the scaling matter era to the epoch of cosmic acceleration. 
Our choice of the double exponential potential does not lose 
the generality for describing such a transition.

\section{Cosmological evolution for 
a concrete dark energy model}\label{Sec:Cosmology}

The model discussed in the previous section does not allow 
for viable cosmological evolution with the scaling 
radiation/matter era followed by the late-time 
dark energy attractor. 
Since the critical point (a) is always 
stable for $\Omega_{\phi}<1$, the scaling solution does not exit to the 
epoch of cosmic acceleration driven by the fixed point (b).
On the contrary, if the parameters are chosen such that
the field has to approach point (b) at late time, 
the scaling behavior at early time is lost.

 In this work, we would like to maintain the scaling behaviour in the early cosmological epoch as this property allows for a natural large value of the energy density of the field in the past despite of its small value at late time.
To realize the proper cosmic expansion history with an early-time scaling period and a late-time dark energy attractor, we construct a model in which the two features 
associated with the critical points (a) and (b)
discussed in Sec.~\ref{Sec:Aconst} are present by adding a second exponential 
potential term similar to that used for Quintessence in Ref.~\cite{Barreiro:1999zs}. 
More precisely, the model has the same $G_3$ function as 
Eq.~(\ref{G3}) but with two exponential potentials of the form:
\be
V=V_1e^{-\beta_1\phi}+V_2e^{-\beta_2\phi}\,,
\label{dpot}
\ee
where $V_1,V_2,\beta_1,\beta_2$ are positive constants
with $\beta_1 \gg {\cal O}(1)$ and $\beta_2 \lesssim {\cal O}(1)$.
The first potential $V_1 e^{-\beta_1 \phi}$ gives rise to 
the scaling fixed point (a) with $\beta=\beta_1$, whereas the second potential 
$V_2 e^{-\beta_2\phi}$ leads to the 
scalar-field dominated point (b) with $\beta=\beta_2$. 
The expansion history of this model is not the same as 
standard Quintessence (hereafter QE) with two potentials 
because it is modified by the $G_3$ term. 
We call this model G3.

We take into account radiation and nonrelativistic matter whose background 
densities are given, respectively, by $\rho_r$ and $\rho_m$, so that 
$\rho_{\gamma}=\rho_r+\rho_m$.
To study the background cosmological dynamics, we define the following 
dimensionless variables:
\ba
& &
y_1=\frac{\sqrt{V_1 e^{-\beta_1\phi}}}{\sqrt{3}H}\,,\qquad 
y_2=\frac{\sqrt{V_2 e^{-\beta_2\phi}}}{\sqrt{3}H}\,, \nonumber \\
& &
\Omega_r=\frac{\rho_r}{3H^2}\,,\qquad 
\Omega_m=\frac{\rho_m}{3H^2}\,,
\ea
with $x=\dot{\phi}/(\sqrt{6}H)$ and $y^2=y_1^2+y_2^2$.
{}From the constraint Eq.~(\ref{Freq}), we obtain 
\be
\Omega_m=1-\Omega_{r}-\Omega_{\phi}\,,
\ee
where 
\be
\Omega_{\phi}=\Omega_{G_2}
+\Omega_{G_3}\,,\qquad 
\Omega_{G_2}=x^2+y_1^2+y_2^2\,,
\ee
with $\Omega_{G_3}$ given by Eq.~(\ref{omeG3}).
We obtain the autonomous equations in the forms:
\ba
x' &=& \frac{1}{\sqrt{6}} \tilde{f}(x,y)
-\frac{\dot{H}}{H^2}x\,,
\label{dxeq}\\
y_i' &=& -\sqrt{\frac32}\beta_i x y_i
-\frac{\dot{H}}{H^2}y_i\,,\\
\Omega_r' &=& -4\Omega_r-2\frac{\dot{H}}{H^2}
\Omega_r\,, 
\label{dOmereq}
\ea
where $i=1,2$. The function $\tilde{f}(x,y)$ follows from Eq.~(\ref{fxy}) 
after the replacements
$\beta y^2 \to \beta_1 y_1^2+\beta_2 y_2^2$ and 
$\gamma \Omega_{\gamma} \to (4/3)\Omega_r
+\Omega_m$, with Eq.~(\ref{gY}).
The derivative term $\dot{H}/H^2$ is given by
Eq.~(\ref{dotH}) with the correspondence
$\gamma \Omega_{\gamma} \to (4/3)\Omega_r+\Omega_m$. 
The dark energy equation of state $w_{\phi}$ follows from 
Eq.~(\ref{wphi}) after the replacements $y^2 \to y_1^2+y_2^2$ and $f \to \tilde{f}$.

The scaling radiation era corresponds to the 
fixed point (a1) given by 
\be
\left( x, y_1, y_2, \Omega_m \right) 
=\left( \frac{2\sqrt{6}}{3\beta_1}, 
\frac{\sqrt{12+6A(3\beta_1-4\lambda)}}{3\beta_1},0,0
\right)\,,
\label{a1}
\ee
and $\Omega_{r}=1-\Omega_{\phi}$ with 
$\Omega_{\phi}=[4+2A(5\beta_1-4\lambda)]/\beta_1^2$,
whereas the scaling matter era is characterized by the 
critical point (a2) given by 
\be
\left( x, y_1, y_2, \Omega_r \right)
=\left( \sqrt{\frac{3}{2}} \frac{1}{\beta_1},
\frac{\sqrt{6+12A(\beta_1-\lambda)}}{2\beta_1},0,0
\right)\,,
\label{a2}
\ee
and $\Omega_{m}=1-\Omega_{\phi}$ with 
$\Omega_{\phi}=3[1+A(3\beta_1-2\lambda)]/\beta_1^2$.
The scalar-field dominated point (b) corresponds to 
\ba
\hspace{-0.6cm}
& &
x=\frac{\beta_2-6 A }{\sqrt{6}[1+A (\beta_2-2\lambda)]}\,,\qquad 
y_1=0\,, \nonumber \\
\hspace{-0.6cm}
& &
y_2=\sqrt{\frac{[1+2A(3A-\lambda)] [6-\beta_2^2
+12 A (\beta_2 -\lambda )]}{6[1+A (\beta_2 -2 \lambda )]^2}},
\label{bx}
\ea
with $\Omega_m=\Omega_r=0$ and $\Omega_{\phi}=1$. 

\begin{figure*}[t]
\centering
\includegraphics[height=3.3in,width=3.5in]{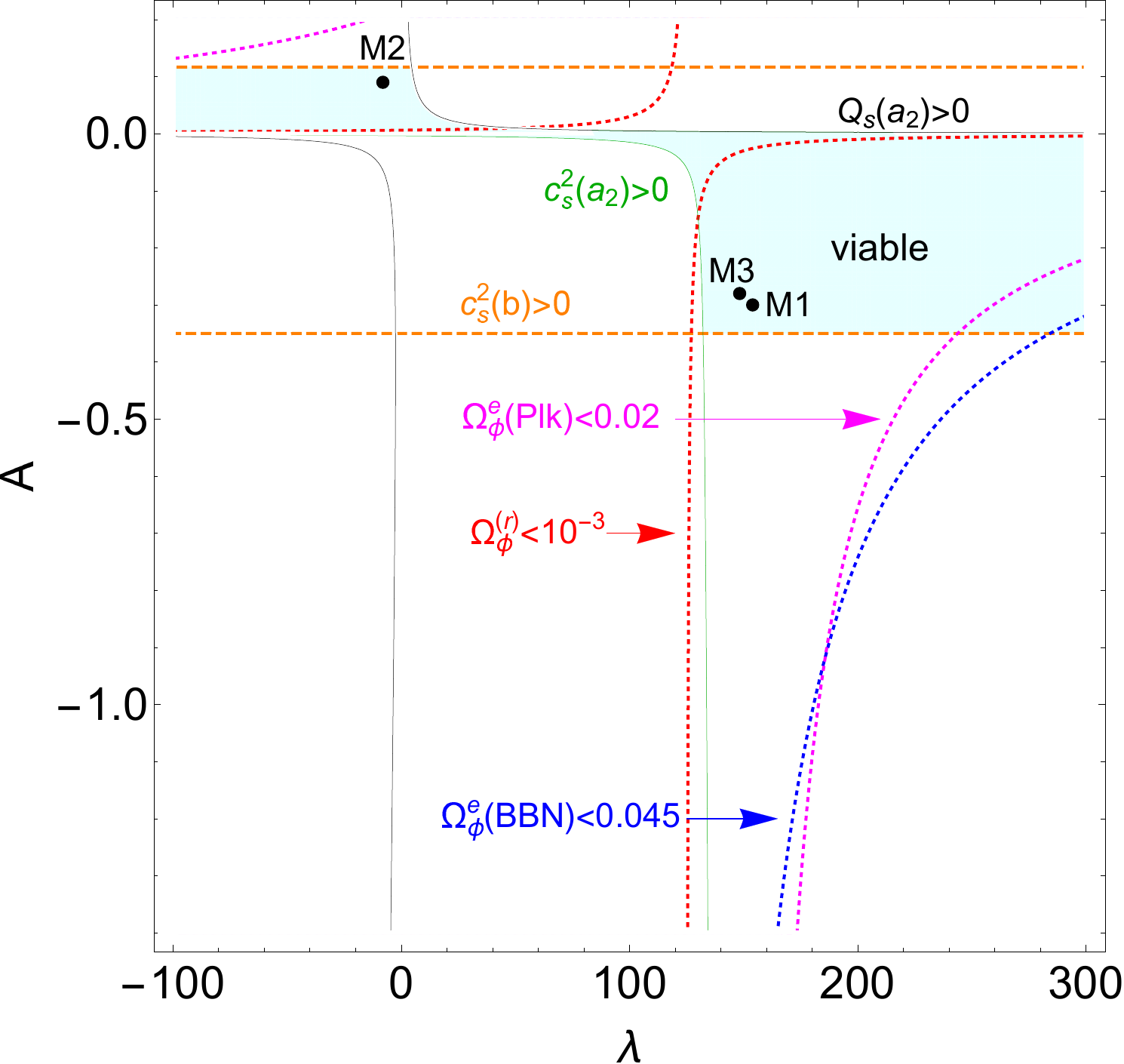} 
\includegraphics[height=3.3in,width=3.5in]{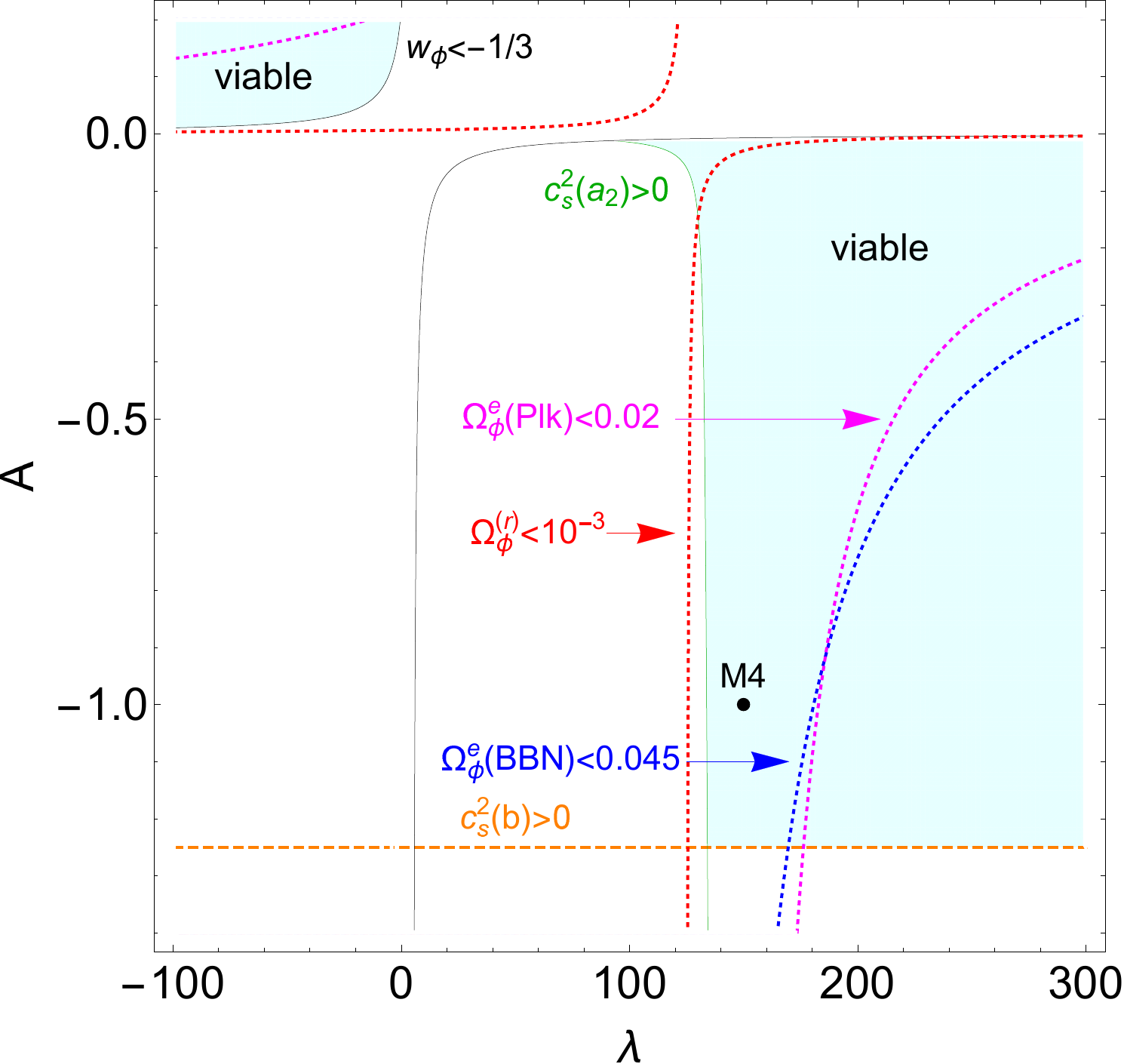}
\caption{\label{stability} 
Viable model parameter spaces (light blue) 
in the $(\lambda, A)$ plane for 
the two cases: (i) $\beta_1=100$, $\beta_2=0.7$ (left) 
and (ii) $\beta_1=100$, $\beta_2=2.5$ (right). 
Each boundary is obtained by using the conditions 
(\ref{acon1})-(\ref{acon3}) for points (a1) and (a2) as well as 
the conditions (\ref{bcon1})-(\ref{bcon6}) 
for point (b). The observational bounds (\ref{earlyDEconst}) 
and (\ref{earlyDEconst2}) are also plotted, together with 
the region $\Omega_{\phi}^{(r)}<10^{-3}$.  
The labels M1, M2, M3, and M4 correspond to the G3 
models presented in Table~\ref{Models}.}
\end{figure*}

Now, we have four parameters $\{\beta_1,\beta_2,\lambda, A\}$ 
in our G3 model.
We choose the two parameters  $\{\beta_1,\beta_2\}$ 
and then constrain the values of $\lambda$ and $A$ 
according to the viability conditions 
discussed in Sec.~\ref{Sec:Aconst}.
The theoretically consistent conditions for points (a1) and 
(a2) are given by Eqs.~(\ref{acon1})-(\ref{acon3}) with 
the replacement $\beta \to \beta_1$, 
where $\gamma=4/3$ for (a1) and $\gamma=1$ for (a2). 
There are also the BBN and CMB bounds (\ref{earlyDEconst}) 
and (\ref{earlyDEconst2}) derived by setting $\beta \to \beta_1$.
For point (b), we also require that the conditions 
(\ref{bcon1})-(\ref{bcon6}) hold with the replacement 
$\beta \to \beta_2$. 
In Fig.~\ref{stability}, we plot the 
allowed parameter space
in the $(\lambda, A)$ plane  (light blue color)  
for (i) $\beta_1=100$, $\beta_2=0.7$ (left) and 
(ii) $\beta_1=100$, $\beta_2=2.5$ 
(right)\footnote{We note that theoretically consistent regions 
resulting from this analysis ensures the viability at critical points 
not along the whole evolution of the system, 
which instead needs to be confirmed.
We have tested for several combinations of parameters in these regions 
and found that the system is stable at any time.}.  

In case (i), the allowed parameter 
space is surrounded by several boundaries determined 
by the conditions $c_s^2 ({\rm a2})>0$, $c_s^2 ({\rm b})>0$, 
$Q_s ({\rm a1})>0$. They translate, respectively, to 
\ba
& &
8A \beta_1-6A (A+\lambda) +3>0\,,\label{Abcon1}\\
& &
-\frac{\beta_2}{2}<A<\frac{\beta_2}{6}\,,\label{Abcon} \\
& &
1+2A \left( 3A-\lambda \right)>0\,,
\ea
where the condition (\ref{Abcon}) corresponds to  
$-0.35<A<0.117$ for $\beta_2=0.7$. 
All the other theoretically consistent conditions
are satisfied in the viable parameter region
plotted in the left panel of Fig.~\ref{stability}. 
The observational bounds (\ref{earlyDEconst}) and 
(\ref{earlyDEconst2}) exclude only a narrow region 
of the theoretically consistent parameter space.  
We also show the parameter space in which the dark energy 
density parameter in the scaling radiation era is in the range
$\Omega_{\phi}^{(r)}<10^{-3}$.
This condition is not obligatory, but we plot such a region  
for the purpose of understanding the parameter space 
in which the primordial scaling value of $\Omega_{\phi}$ 
is small.

In case (ii), in the limit $A \to 0$, 
the slope $\beta_2=2.5$ is too large to satisfy 
the stability condition (\ref{bcon5}) of point (b). 
Moreover, for $A=0$, the cosmic acceleration occurs 
for $w_{\rm eff}=w_{\phi}=-1+\beta_2^2/3<-1/3$, i.e., 
$\beta_2^2<2$.
On the other hand, the nonvanishing cubic coupling $A$ 
allows for the possibility of cosmic acceleration 
even for $\beta_2^2>2$.
Indeed, the viable region for $A>0$ 
is determined by the condition 
$w_{\rm eff}({\rm b})=w_{\phi}({\rm b})<-1/3$, i.e., 
\be
\frac{\beta_2 (\beta_2-6A)}
{1+A(\beta_2-2\lambda)}-2<0\,,
\label{adcon}
\ee
under which $\mu_1({\rm b})<0$.
Note that the condition (\ref{adcon}) also determines 
the upper border for $A<0$ (see the right panel 
of Fig.~\ref{stability}).
The other boundaries of viable parameter space are 
determined by the conditions $c_s^2 ({\rm a2})>0$ and 
$c_s^2 ({\rm b})>0$, i.e., 
by Eqs.~(\ref{Abcon1}) and (\ref{Abcon}). 
The important difference from case (i) is that the observational 
bounds (\ref{earlyDEconst}) and (\ref{earlyDEconst2}) restrict
a broader range of theoretically consistent model parameters.
Since there exists a viable parameter space even for $\beta_2^2>2$, 
the cubic coupling allows a wider allowed range of $\beta_2$ 
compared to QE.

\begin{table}[t!]
\begin{center}
\begin{tabular}{|c|c|c|c|c|c|c|c|}
\hline
Model & $\beta_2$ & $A$ & $ \lambda$ & 
$w_{\phi}^{(0)}$ & $w_{\rm eff}^{(0)}$ & 
$\Omega_\phi ( z=50)$\\ 
\hline\hline
M1  &0.7 & -0.3 & 154  & -0.993 & -0.675 & $1.0 \times 10^{-3}$ 
\\ \hline 
M2  &0.7 & 0.09 & -8   &  -0.988 & -0.672 & $8.9 \times 10^{-3}$ 
\\ \hline 
M3  &0.7 &-0.28 & 148.3& -0.993 &  -0.675 &
$4.3 \times 10^{-5}$ \\ \hline
M4  &2.5 & -1   &  150 & -0.975 &  -0.663 &
$3.6 \times 10^{-4}$
\\   \hline
QE1  &0.7 & 0    &   0  &-0.927 & -0.630 & $3.2 \times 10^{-4}$ \\ \hline 
QE2 &2.5 & 0    & 0    & -0.358 & -0.167 & $3.3 \times 10^{-4}$ \\ \hline 
\end{tabular}
\end{center}
\caption{\label{Models} 
Model parameters $\beta_2, A, \lambda$ used 
in the numerical simulations of 
Figs.~\ref{M1densityics}-\ref{wDE}. 
For all of them, $\beta_1=100$. 
In each model, we also show today's dark energy equation 
of state $w_{\phi}^{(0)}$, today's effective equation of 
state $w_{\rm eff}^{(0)}$, and the dark energy 
density parameter $\Omega_\phi$ at the redshift 
$z=50$. Except for QE2, all the other models give rise 
to the cosmic acceleration today 
($w_{\rm eff}^{(0)}<-1/3$).}
\end{table}

\begin{figure*}[t!]
\centering
\includegraphics[height=2.8in,width=3.4in]{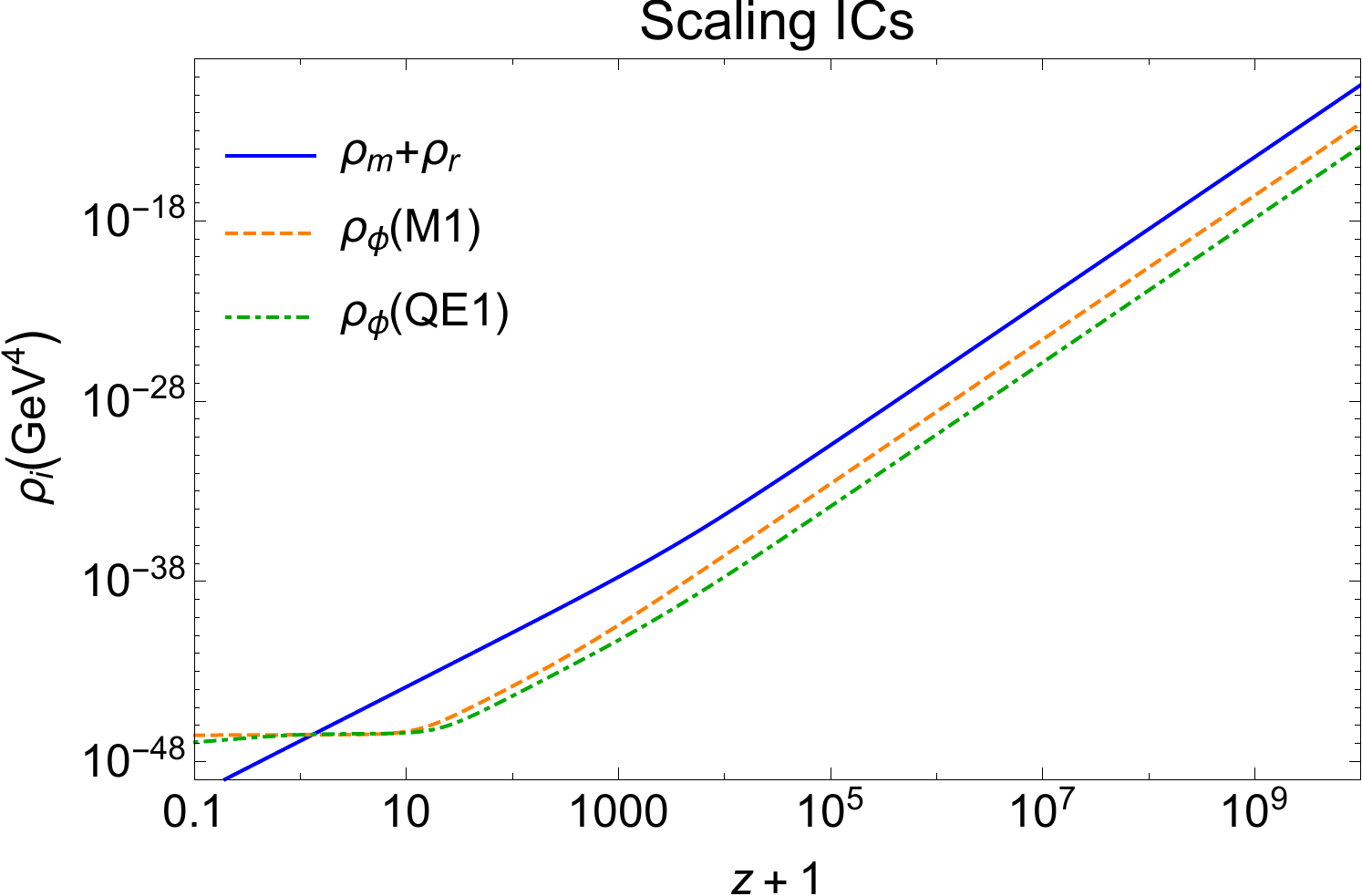}
\includegraphics[height=2.8in,width=3.4in]{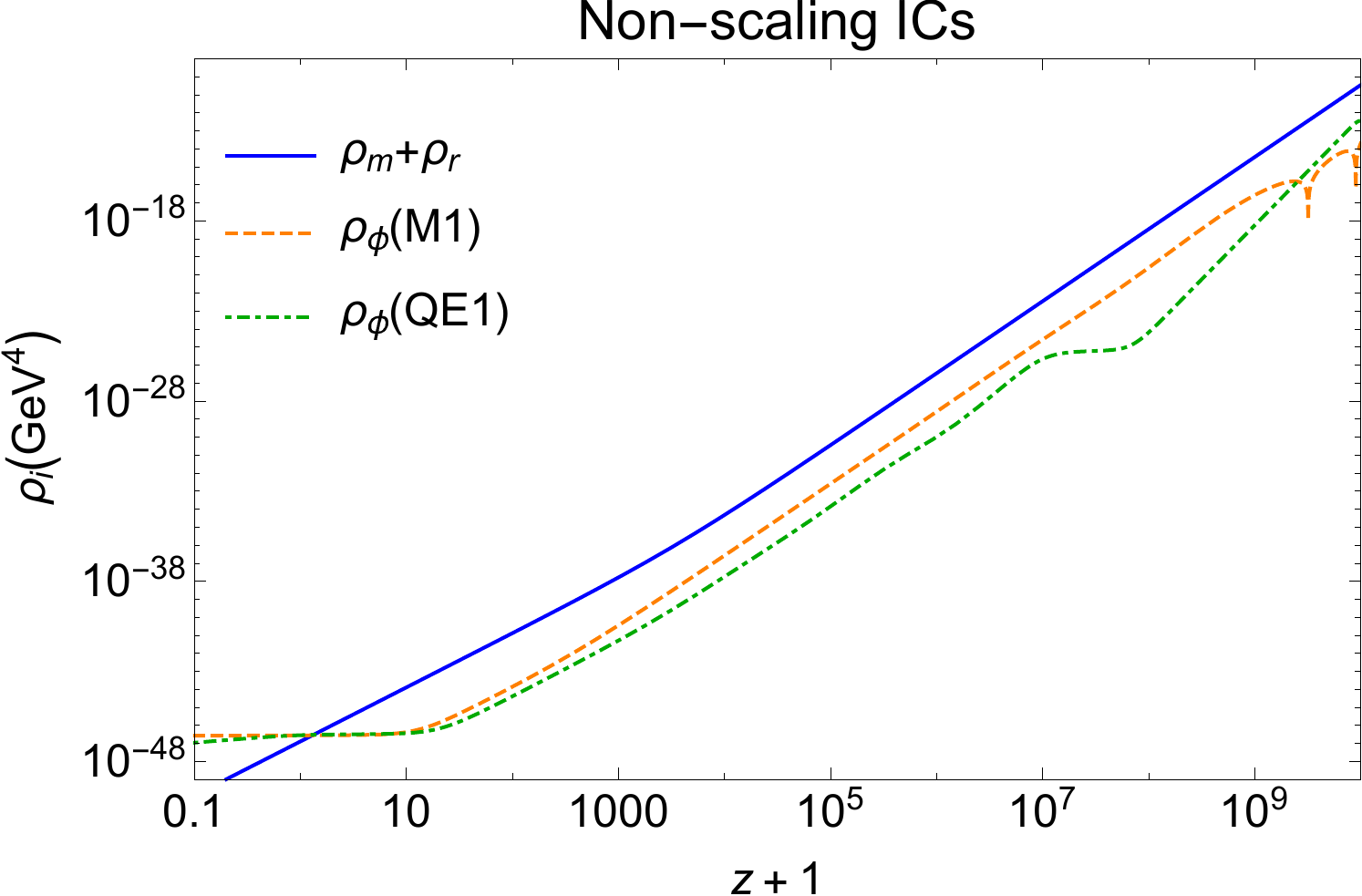}
 \caption{\label{M1densityics} 
(Left) Evolution of the total fluid density
$\rho_m+\rho_r$ (blue, solid line) and the scalar-field 
density $\rho_{\phi}$ for M1 (orange, dashed line) and 
QE1 (green, dot-dashed line). 
The model parameters for M1 and QE1 are 
given in Table \ref{Models}.
The ICs of $x, y_1$ are chosen to be close to those of 
critical point (a), with $y_2, \Omega_r$ 
realizing today's density parameters 
$\Omega_{\phi}^{(0)}=0.68$ and 
$\Omega_r^{(0)} =10^{-4}$.
(Right) Evolution of $\rho_m+\rho_r$ and $\rho_{\phi}$ 
for the same model parameters as those in the left, 
but with different ICs: 
$x=0.015$, $y_1=0.04$ for M1 and 
$x=0.015$, $y_1=0.1$ for QE1. }
\end{figure*}

In Table \ref{Models}, we show four different models M1, M2, M3, and M4, 
all of which are inside the viable region depicted in Fig.~\ref{stability}. 
We also consider two QE models with $\beta_1=100$: 
QE1 ($\beta_2=0.7$) and QE2 ($\beta_2=2.5$).
In the following, we study the cosmological evolution in 
these models by paying particular attention to the effect of the
cubic coupling on the background dynamics.
In doing so, we first comment on the issue of ICs. 
Unless otherwise stated, we select the ICs of $x$ and $y_1$ 
corresponding to the critical point (a1), 
which fix the background dynamics.
The ICs of $y_2$ and $\Omega_r$ are chosen such that today's 
density parameters of $\phi$ and radiation are   
$\Omega_\phi^{(0)}=0.68$ and 
$\Omega_r^{(0)}=10^{-4}$, respectively.
For QE2, the scalar field does not give rise to the 
 late-time cosmic acceleration, in which case 
we identity the present epoch by the condition 
$\Omega_r^{(0)}=10^{-4}$.
We start integrating Eqs.~(\ref{dxeq})-(\ref{dOmereq}) from 
the initial redshift $z_i=10^{10}$.
We also discuss the case in which the ICs of 
$x$ and $y_1$ deviate from point (a1). 

In Fig.~\ref{M1densityics}, the evolution of 
scalar-field density $\rho_{\phi}$ for M1 and QE1 
is plotted, together with the total fluid density 
$\rho_m+\rho_r$. 
In the left panel, the ICs of $x$ and $y_1$ are identical 
to those of point (a1). 
Indeed, the scalar field exhibits scaling behavior 
with the background fluid in the early cosmological epoch
($\rho_{\phi} \propto \rho_m+\rho_r$).
In this case, the field density parameters corresponding to 
points (a1) and (a2) are given by 
$\Omega_{\phi} ({\rm a1})=7.4 \times 10^{-3}$ and 
$\Omega_{\phi} ({\rm a2})=1.9 \times 10^{-3}$, respectively, 
which are consistent with the bounds (\ref{earlyDEconst}) and (\ref{earlyDEconst2}). 
They are by one order of magnitude larger than the corresponding 
values in QE, i.e.,
$\Omega_{\phi} ({\rm a1})=4.0 \times 10^{-4}$ and 
$\Omega_{\phi} ({\rm a2})=3.0 \times 10^{-4}$. 
Indeed, this property can be confirmed in the left panel of 
Fig.~\ref{M1densityics}. 

In the right panel of Fig.~\ref{M1densityics}, we show the 
evolution of $\rho_{\phi}$ and $\rho_m+\rho_r$ 
for M1 by changing the ICs of $x$ by $0.01\%$ 
and $y_1$ by $1\%$. 
At the same time, the ICs for QE1 are changed  
by 0.01\% in $x$ and by 10\% in $y_1$.
For M1, the solutions approach the scaling critical point 
(a1) after a few oscillations in the field density. 
For QE1, the larger change of $y$ relative to M1 
does not induce oscillations in the field density, 
but its takes some time to reach the scaling regime. 
The important point is that, even in presence 
of the cubic coupling $G_3$, the first exponential potential 
$V_1e^{-\beta_1 \phi}$ leads to stable scaling fixed points 
(a1) and (a2) with $\Omega_{\phi}<1$ that always 
attracts solutions with different ICs.

Since the additional exponential potential 
$V_2e^{-\beta_2 \phi}$ 
is present, the solutions finally exit from the scaling matter era 
to the epoch of cosmic acceleration driven by the critical 
point (b). 
{}From Eqs.~(\ref{wphib}) and (\ref{OmeG3b}), the dark energy equation 
of state and the density parameter  arising from $G_3$ at point (b) are 
\ba
w_{\phi} &=&
-1+\frac{\beta_2 (\beta_2 -6 A)}
{3[1+A(\beta_2-2\lambda)]}\,,
\label{wphiG3}\\
\Omega_{G_3} &=&
\frac{A(\beta_2-6A)[6+6A (\beta_2-\lambda)-\beta_2 \lambda]}
{3[1+A(\beta_2-2\lambda)]^2}\,,
\label{OmeG32}
\ea
which give $w_{\phi}=-0.994$ and 
$\Omega_{G_3}=-5.0 \times 10^{-3}$ for M1.
The density associated with the $G_3$ term is suppressed 
at low redshifts, so that the dominant contribution to 
$\Omega_{\phi}$ comes from the standard field
density $\Omega_{G_2}$.
For QE1 we have $w_{\phi}=-0.837$ at point (b), 
so the field density 
$\rho_{\phi}$ for M1 decreases more slowly relative to 
that for QE1 in the future ($z<0$). 
This behavior can be confirmed in the numerical 
simulation of Fig.~\ref{M1densityics}.

\begin{figure}[t!]
\centering
\includegraphics[height=2.6in,width=3.1in]{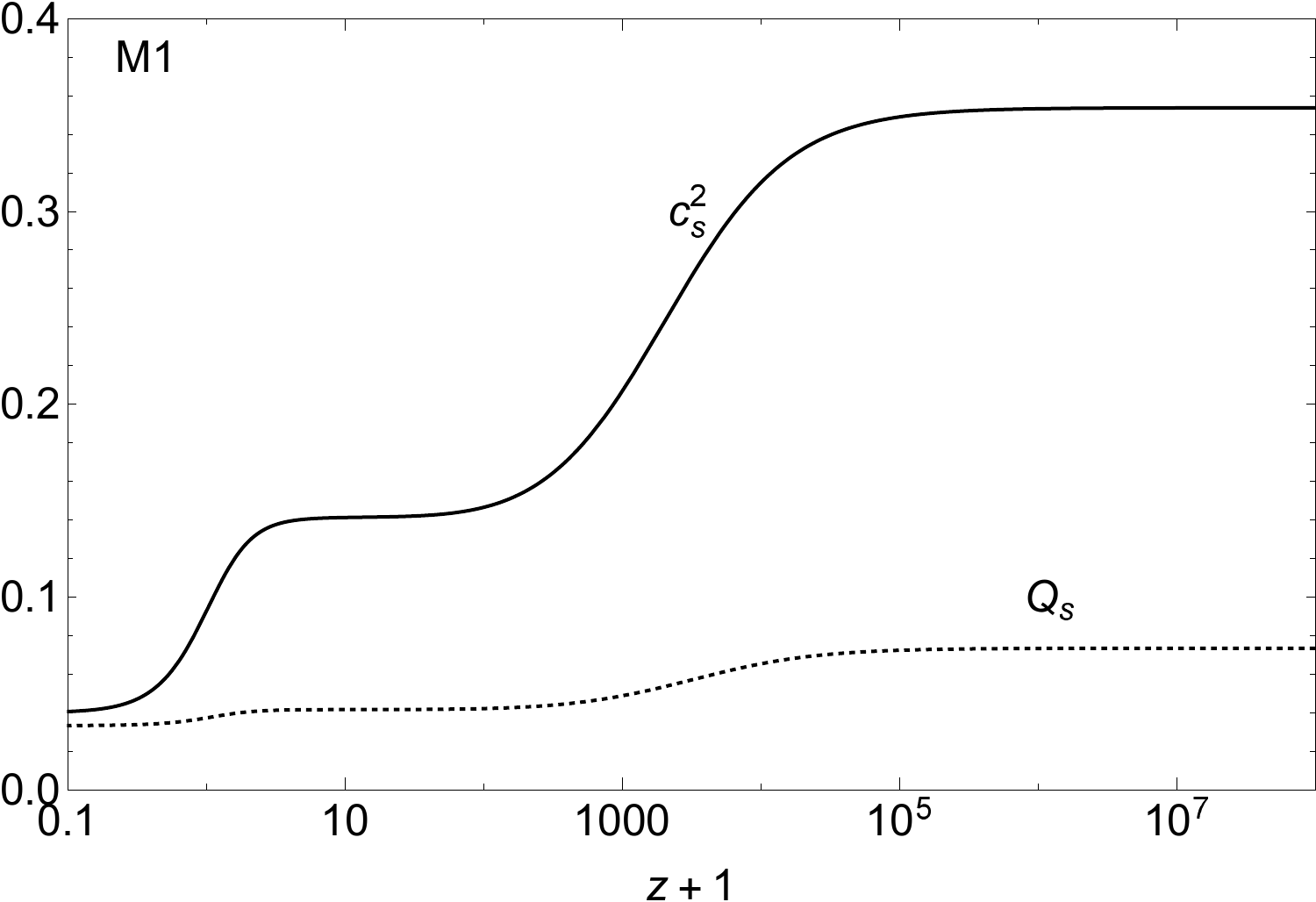}
\caption{\label{csQs} Evolution of the propagation speed 
squared $c_s^2$ (solid line) and the kinetic term $Q_s$ 
(dotted line) versus $z+1$ for the model M1.  }
\end{figure}

To ensure the absence of ghost and Laplacian instabilities, 
we need to confirm that $Q_s$ and $c_s^2$ 
given by Eq.~(\ref{stabAconst}) and (\ref{stabAconst2}) 
remain positive.
In Fig.~\ref{csQs}, we plot the evolution of those 
quantities for the model M1 by choosing ICs same as those 
used in the left panel of Fig.~\ref{M1densityics}.
The values of $c_s^2$ on points (a1) and (a2) can be obtained 
by substituting $\gamma=4/3$ and $\gamma=1$ with 
$\beta=\beta_1$ into $c_s^2$ 
at point (a) given in Table \ref{critp}, respectively, 
while $c_s^2=1+8A/(\beta_2-6A)$ on point (b). 
They are in good agreement with the numerical 
simulation of Fig.~\ref{csQs}. 
Moreover, during the transient regimes between 
critical points, $c_s^2$ remains positive without crossing 0. 
This is also the case for $Q_s$, so the model M1 suffers 
neither ghost nor Laplacian instabilities during the whole 
cosmological evolution.
We have confirmed that such conditions are also satisfied for 
all the models listed in Table~\ref{Models}.

\begin{figure}[t]
\centering
\includegraphics[height=2.5in,width=3.3in]{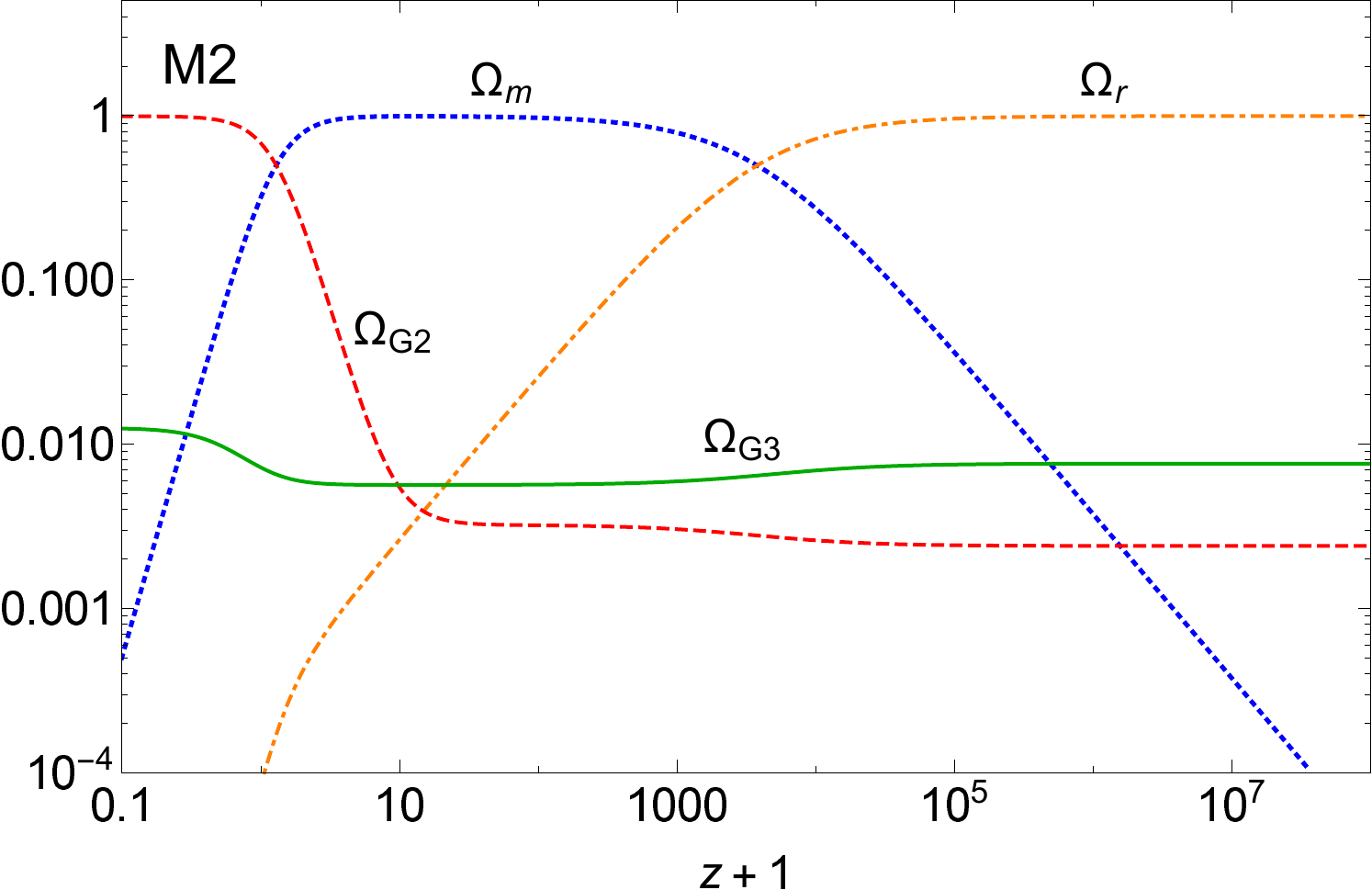}
\caption{\label{DensityM2} 
Evolution of $\Omega_m$ (blue, dotted line), $\Omega_r$ (orange, dot-dashed line), 
$\Omega_{G_2}$ (red dashed line) and $\Omega_{G_3}$ (green solid line) 
versus $z+1$ for the model M2. 
Note that $\Omega_{G_2}$ and $\Omega_{G_3}$ are the 
density parameters arising from the field Lagrangians 
$G_2$ and $-G_3 \square \phi$, respectively. }
\end{figure}

In Fig.~\ref{DensityM2}, the evolution of density parameters is plotted for the model M2, 
which exists in the region 
$\lambda<0$ and $A>0$ in the left panel 
of Fig.~\ref{stability}.
We observe that $\Omega_{G_3}$ dominates over 
$\Omega_{G_2}$ during the early cosmological epoch, 
but the main contribution to $\Omega_{\phi}$ 
comes from $\Omega_{G_2}$ at redshifts 
$z \lesssim 10$.
{}From Eq.~(\ref{OmeG32}), we have 
$\Omega_{G_3}=0.012$ on point (b), 
which is positive. This property is 
different from the model M1, in which 
$\Omega_{G_3}$ is negative on point (b). 
The important point is that the cubic coupling can provide
the dominant contribution to $\Omega_{\phi}$ 
in the scaling radiation and matter eras, but its effect 
on $\Omega_{\phi}$ tends to be suppressed ($|\Omega_{G_3}| \ll 1$) 
at low redshifts.

{}From Eq.~(\ref{wphiG3}), the dark energy equation of state 
on point (b) for M2 is given by $w_{\phi}=-0.985$, which 
is again closer to $-1$ relative to the value $-0.837$ for QE1.
Indeed, it approaches the value $w_{\phi}=-1$ as 
the model shifts to the upper boundary $A=\beta_2/6$ 
of the viable region plotted in the left panel 
of Fig.~\ref{stability}. At the same time, 
the contribution of $\Omega_{G_3}$ to $\Omega_{\phi}$
decreases toward 0. For some specific 
models like covariant Galileons, the field density  
dominated by cubic interactions at low redshifts can give rise to 
the galaxy-ISW anti-correlation incompatible with current 
observations \cite{Kimura:2011td,Renk}, so it is anticipated that the G3 models
satisfying the condition $|\Omega_{G_3}| \ll 1$ at late times
may evade such constraints.

\begin{figure}[t]
\centering
\includegraphics[height=2.4in,width=3.3in]{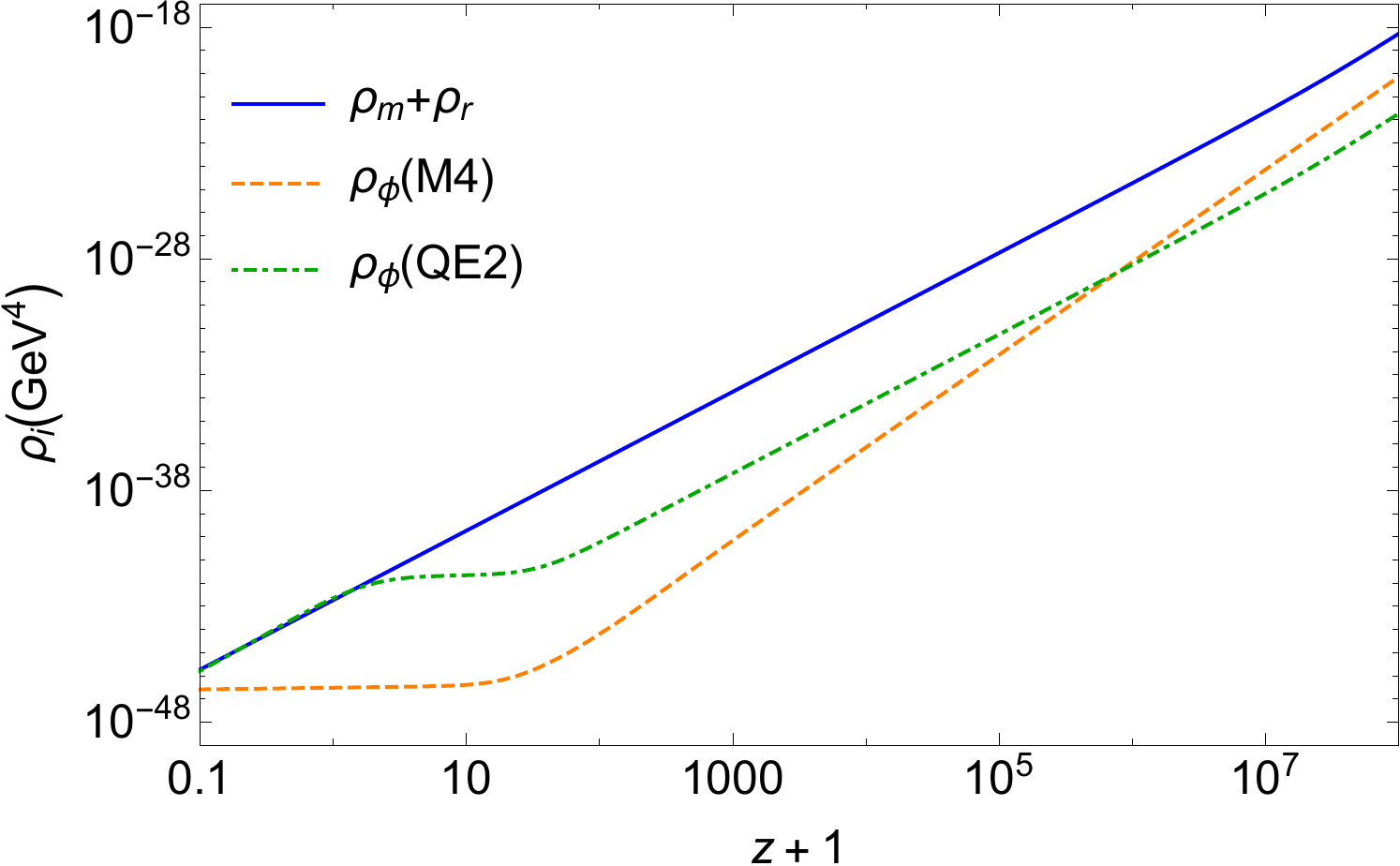}
\caption{\label{DensityM4} 
Evolution of $\rho_{\phi}$ for the model M4 (orange, 
dashed line) and for QE2 (green, dot-dashed line) versus $z+1$, together 
with the background matter density $\rho_m+\rho_r$ (blue, solid line). }
\end{figure}

As we see in the right panel of Fig.~\ref{stability}, there 
are models in which all the theoretically consistent conditions 
are satisfied even for $\beta_2^2>2$.
In Fig.~\ref{DensityM4}, we show the evolution of 
$\rho_{\phi}$ and $\rho_m+\rho_r$ for the models 
M4 and QE2 (in which $\beta_2=2.5$). 
For QE2, the eigenvalue 
$\mu_1$ on point (b) is positive with 
$w_{\rm eff}=w_{\phi}>-1/3$, so the Universe does 
not enter the stage of cosmic acceleration.
Instead, the scaling matter era 
($\Omega_{\phi}=3/\beta_1^2=3.0 \times 10^{-4}$) is 
followed by the other scaling matter 
fixed point (a2) driven by the second exponential potential 
$V_2e^{-\beta_2 \phi}$ with 
$\Omega_{\phi}=3/\beta_2^2=0.48$.
As we see in Table \ref{Models}, today's value of 
$w_{\rm eff}^{(0)}$ for QE2 is larger than $-1/3$, so
the Universe does not exhibit the cosmic 
acceleration today.

For the model M4, the coupling $G_3$ allows the possibility 
for realizing the scaling radiation era with $\Omega_{\phi}=0.02$.
While $\Omega_{G_3}$ dominates over $\Omega_{G_2}$ 
on point (a1), the contribution $\Omega_{G_3}$ exactly 
vanishes on point (a2) for M4 and hence 
$\Omega_{\phi}=3/\beta_1^2=3 \times 10^{-4}$ . 
Since $\Omega_{\phi}$ for point (a2) is by two 
orders of magnitude smaller than that for point (a1), 
it takes some time for the solutions to move from (a1) to (a2). 
Indeed, the second exponential potential 
$V_2 e^{-\beta_2 \phi}$ starts to contribute to the field density 
before the 
solutions completely approach point (a2). 
This is the reason why $\rho_{\phi}({\rm M4})$ in 
Fig.~\ref{DensityM4} decreases 
faster than $\rho_m+\rho_r$ in the redshift range  
$10 \lesssim z \lesssim 1000$.

For M4, around the redshift $z \lesssim 10$, 
the solutions start to approach point (b) 
characterized by $w_{\phi}=-0.976$ and 
$\Omega_{G_3}=-0.016$, so the cosmic acceleration 
occurs even for $\beta_2^2>2$.
In spite of the dominance of $\Omega_{G_3}$ in 
the scaling radiation era, $\Omega_{G_3}$ is suppressed relative 
to $\Omega_{G_2}$ at low redshifts.
With this example, we showed that the allowed parameter 
space for $\beta_2$ is wider than that for QE.

\begin{figure}[t]
\centering
\includegraphics[height=2.5in,width=3.3in]{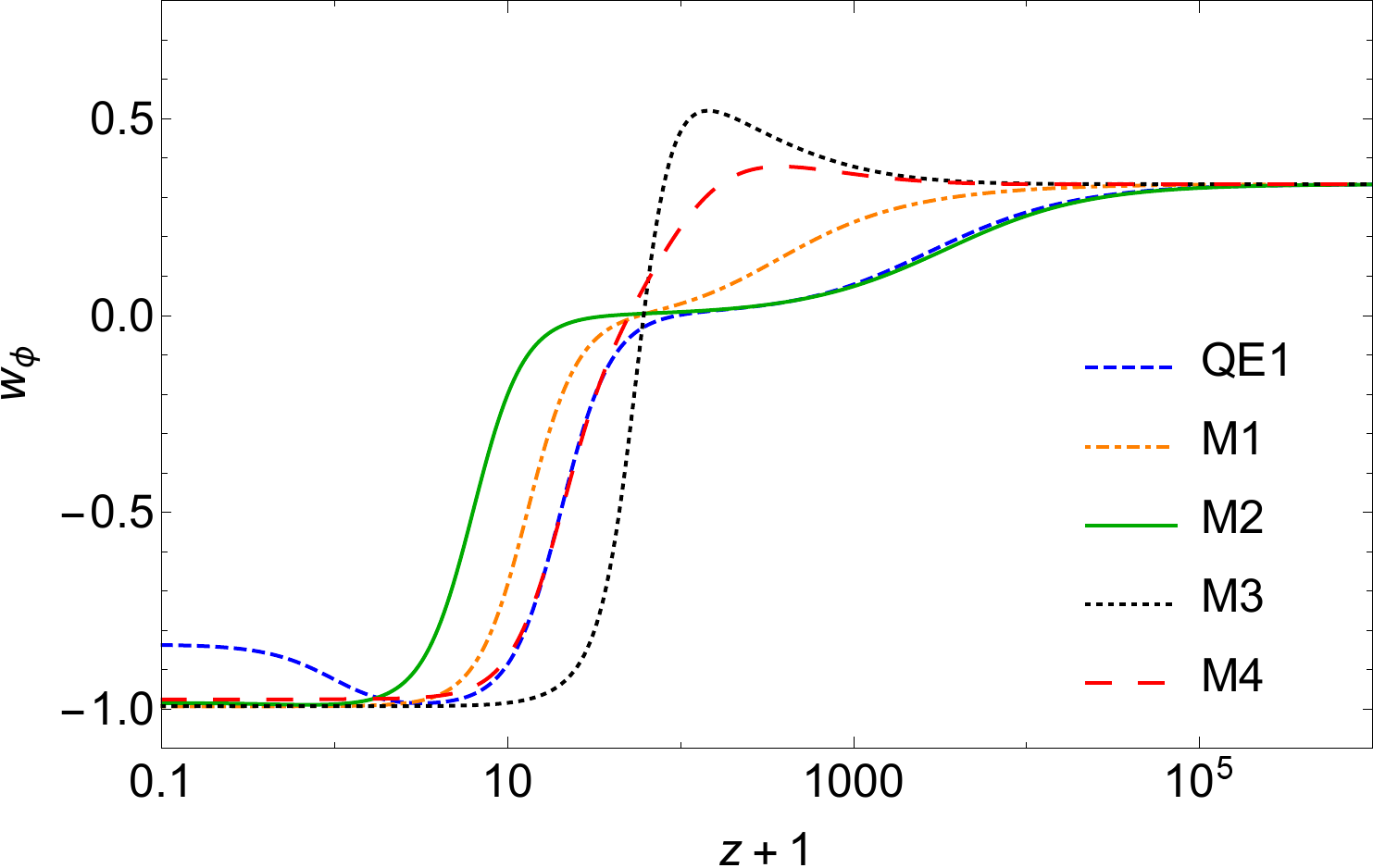}
\caption{\label{wDE} 
Variation of $w_\phi$ versus $z+1$ for the models: 
QE1 (blue dashed line), M1 (dot-dashed orange line), 
M2 (green solid line), M3 (dotted black line), 
and M4 (long dashed red line). 
 }
\end{figure}

In Fig.~\ref{wDE}, we plot the evolution of $w_{\phi}$ for 
all the G3 models and QE1 presented in Table \ref{Models}. 
For M1 and M2, the scaling radiation era ($w_{\phi} \simeq 1/3$) 
is followed by the scaling matter epoch ($w_\phi \simeq 0$).
In the models M3 and M4, the scaling matter era is practically absent 
by reflecting the fact that $\Omega_{\phi}$ at point (a2) is 
much smaller than that at point (a1), e.g., 
$\Omega_{\phi}({\rm a1})=5.6 \times 10^{-3}$ and 
$\Omega_{\phi}({\rm a2})=1.4 \times 10^{-5}$ for M3.
In such cases, the matter-dominated epoch corresponds 
to the transient period between critical points (a1) and (b). 
For M3, the field density parameter at the redshift $z=50$
is $\Omega_{\phi}=4.3 \times 10^{-5}$, which is 
smallest among the G3 models studied above.
The Planck bound (\ref{earlyDEconst2}) is satisfied for all the 
models listed in Table \ref{Models}.

The dark energy equation of state 
today ($w_{\phi}^{(0)}$) is related to the model parameters  
$A$ and $\beta_2$. For increasing $|A|$ from 0, 
we need to choose
larger values of $\beta_2$ for the reason of theoretical viability.
The larger $\beta_2$ results in $w_{\phi}^{(0)}$ deviating from $-1$. 
In Fig.~\ref{wDE}, we observe that M4 gives the highest 
values of $w_{\phi}^{(0)}$ and $w_{\rm eff}^{(0)}$
among the G3 models listed 
in Table \ref{Models}. 
However, all the G3 models give rise to $w_{\phi}^{(0)}$ closer to 
$-1$ than that in QE1. This behavior of G3 models is also in better agreement 
with the latest cosmological constraints \cite{Ade:2015xua,Troxel:2017xyo,Abbott:2017wau} 
compared to QE with the large deviation of 
$w_{\phi}^{(0)}$ from $-1$.

In summary, we found the following new features in our 
cubic Horndeski model.
\begin{itemize}
\item The model allows for a scaling behavior at early time and 
a dark energy attractor at late time;
\item The viable parameter space is wider than that for QE;   
\item There exist  G3 models in which the dark energy equation of state today ($w_{\phi}^{(0)}$) 
is close to  $-1$ even for $\beta_2^2>2$;
\item The model can be consistent with the BBN and CMB bounds 
(\ref{earlyDEconst}) and (\ref{earlyDEconst2});
\item The cubic coupling can provide the dominant contribution to $\Omega_{\phi}$ in the early scaling
epoch, but its contribution to the field density is typically suppressed at low redshifts 
($|\Omega_{G_3}| \ll \Omega_{G_2}$).
\end{itemize}
%

\section{Insight on modifications of gravity 
at large scales}
\label{persec}

Finally, we discuss the impact of our G3 models on the 
evolution of linear scalar perturbations relevant to the 
growth of structures. 
Let us consider the perturbed line element on 
the flat FLRW background given by 
\be
ds^2=-\left( 1+2\Psi \right) dt^2
+a^2(t) \left(1-2\Phi \right) \delta_{ij}dx^i dx^j\,,
\ee
where $\Psi$ and $\Phi$ are gravitational potentials 
in the Newtonian gauge. 
We define the gravitational slip parameter 
\be
\eta=\frac{\Phi}{\Psi}\,,
\ee
which characterizes the difference between two 
gravitational potentials.

For the matter sector, we take into account nonrelativistic 
matter components with the background density 
$\rho_i$ and the density contrast 
$\Delta_i=\delta \rho_i/\rho_i$. 
In Fourier space with the coming wavenumber $k$, 
we relate $\Psi$ and the total matter density perturbation 
$\rho \Delta=\sum_{i} \rho_i \Delta_i$, 
as \cite{Bean:2010zq,Silvestri:2013ne}
\be
-k^2\Psi=4\pi G_N a^2\mu\,\rho \Delta\,,
\label{mudef}
\ee
which corresponds to the modified Poisson equation. 
If the quantity $\mu$ deviates from 1, this leads to 
the modified growth of matter density contrast $\Delta$ 
compared to that in GR.
Along with $\mu$ and $\eta$, we also define a quantity 
$\Sigma$ that relates the weak lensing potential $\Psi+\Phi$ with $\Delta$, as 
\be
-k^2\left(\Psi+\Phi \right)=8\pi Ga^2\Sigma \rho\Delta\,,
\label{sigmadef}
\ee
where 
\be
\Sigma=\frac{1+\eta}{2}\mu\,.
\ee
In GR with the field Lagrangian $G_2(\phi, X)$  we have $\eta=\mu=\Sigma=1$, 
so any departure from these values translates to a signature of the modification of gravity.

\begin{figure}[h!]
\centering
\includegraphics[height=2.5in,width=3.3in]{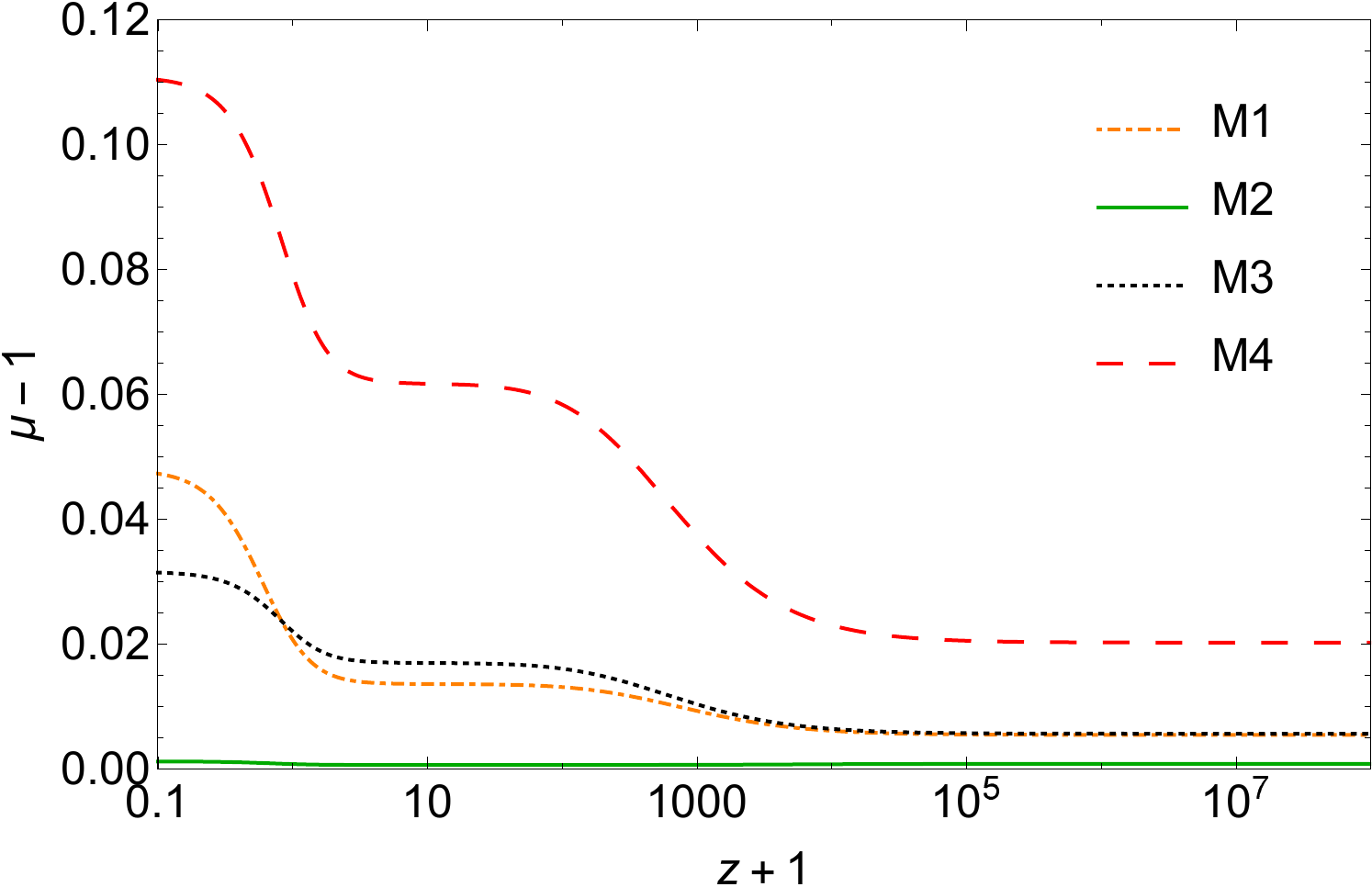}
\caption{\label{fig:mu} 
Evolution of the quantity $\mu-1$ versus $z+1$ for the 
models M1 (orange, dot-dashed line), 
M2 (green, solid line), M3 (black, dotted line), 
and M4 (red, long dashed line). }
\end{figure}

For perturbations relevant to the observations of 
large-scale structures and weak lensing, we are 
interested in the modes deep inside the sound horizon 
($c_s^2 k^2 \gg a^2 H^2$).
Provided that the oscillating mode of scalar-field perturbations 
is negligible compared to the matter-induced mode, we can 
resort to the so-called quasi-static approximation under which 
the dominant contributions to the perturbation equations of 
motion correspond to those containing $k^2/a^2$ 
and $\delta \rho_i$ \cite{Boi,DKT}. 
In our cubic Horndeski theory, the quasi-static approximation 
for the modes deep inside the sound horizon 
gives\footnote{In Ref.~\cite{Kase:2018iwp}, 
these results were derived by 
choosing the unitary gauge in which the field perturbation 
$\delta \phi$ vanishes, but $\eta, \mu, \Sigma$ 
are gauge-invariant quantities in the small-scale limit. 
} \cite{Kase:2018iwp}
\be
\eta=1\,,\qquad
\mu=\Sigma= 1+\frac{\alpha_{\rm B}^2}
{Q_s c_s^2(1+\alpha_{\rm B})^2}\,,
\label{mure}
\ee
where the braiding parameter 
$\alpha_{\rm B}$ \cite{Bellini:2014fua} is given by 
\be
\alpha_{\rm B}
=-\frac{\dot{\phi}XG_{3,X}}{H}
=-\sqrt{6}Ax \,.
\ee
On using Eqs.~(\ref{stabAconst}) and (\ref{stabAconst2}), 
it follows that 
\be
\mu=\Sigma=1+\frac{6A^2 x}
{4\sqrt{6}A-3x (2A^2+2A \lambda-1)}\,.
\label{mucon}
\ee
{}From the above expressions there is no gravitational 
slip ($\Psi=\Phi$), but the cubic coupling $G_3$ modifies the growth 
of structures ($\mu\neq 1$) and the evolution of 
weak lensing potential ($\Sigma\neq 1$). 
The deviation of $\Sigma$ from 1 also gives rise to  
modifications to the integrated Sachs-Wolfe effect in the CMB. 
Moreover, we infer from Eq.~(\ref{mure}) that 
$\mu=\Sigma>1$ 
under the absence of scalar ghosts ($Q_s>0$) and 
Laplacian instabilities ($c_s^2>0$).

On the critical points (a1), (a2), and (b) discussed in 
Sec.~\ref{Sec:Cosmology}, the quantity $\mu$ 
reduces to 
\ba
{\rm (a1)}~~
\mu &=& 1+\frac{2A^2}{2A\beta_1+1-2A(A+\lambda)}\,,
\label{muab0}\\
{\rm (a2)}~~
\mu &=& 1+\frac{6A^2}{8A \beta_1+3-6A (A+\lambda)}\,,\\
{\rm (b)}~~
\mu &=& 1+\frac{2A^2 (\beta_2-6A)}
{(\beta_2+2A)[1+2A(3A-\lambda)]}\,.
\label{muab}
\ea
In the scaling radiation and matter epochs driven by 
points (a1) and (a2), respectively, $\mu$ depends 
on $A, \lambda, \beta_1$. On the scalar-field dominated 
point (b), $\mu$ is affected by $\beta_2$ besides $A, \lambda$.

In Fig.~\ref{fig:mu} we plot the evolution of $\mu-1$ for all the G3 models listed in Table \ref{Models}. 
Around the critical points (a1), (a2), (b), the numerical 
values of $\mu-1$ exhibit good agreement with the 
analytic results (\ref{muab0})-(\ref{muab}), e.g.,  
in the model M1, $\mu-1=0.0054, 0.014, 0.048$ 
for (a1), (a2), (b), respectively. 
The quantity $\mu-1$ increases with time for all the cases 
shown in Fig.~\ref{fig:mu}. 
The deviation of $\mu$ from 1 today is about 
$2\%$ for M1 and M3,  $0.08\%$ for M2, and 
$8\%$ for M4. As expected from Eqs.~(\ref{muab0})-(\ref{muab}), 
the largest deviation from GR arises for the model 
with highest value of $A^2$. 

In Ref.~\cite{Peirone:2017ywi}, the authors studied  phenomenological constraints 
on $\mu$ and $\Sigma$ by using a specific parametrization for time-dependent 
functions (Pad\'e functions) appearing in the effective field theory of dark energy. 
The parameter space exists mostly in the region $(\mu-1)(\Sigma-1)\geq 0$.
In particular, if the stability conditions (absence of ghost and 
Laplacian instabilities) and observational priors are imposed, 
it was shown that the region with $\mu \geq 1$ and $\Sigma \geq 1$ is 
most favored. Since our G3 models predict 
$\mu=\Sigma>1$, they are consistent with the recent 
bounds on $\mu$ and $\Sigma$.

\section{Conclusions} \label{Sec:Conclusions}

In this paper, we have proposed a viable model of cosmic 
acceleration in the framework of cubic-order Horndeski theories. 
We searched for scaling solutions to alleviate the coincidence problem 
of dark energy in the presence of cubic Horndeski interactions besides
the standard field kinetic term with two exponential potentials. 
Extending the analysis of scaling solutions performed 
for the K-essence Lagrangian 
$G_2(\phi, X)$ \cite{Piazza:2004df,Tsujikawa:2004dp}, we assumed that the cubic coupling 
$G_3$ is a function of $Y=Xe^{\lambda\phi}$ and 
found a new type of scaling solutions for the  
coupling $G_3(Y)=A \ln Y$.

In Sec.~\ref{Sec:Aconst}, we first performed a thorough
dynamical analysis of the background cosmology 
for the cubic coupling $G_3(Y)=A\ln Y$ with 
a single exponential potential $V(\phi)=V_0e^{-\beta \phi}$. 
We derived critical points of the system and studied their stability. 
For each critical point, we also discussed 
conditions for the absence of ghost and Laplacian instabilities 
in the small-scale limit. 
The scaling solution (a) and the scalar-field dominated solution (b) 
given in Table \ref{critp} are the two important 
critical points of our system. 
On using theoretically consistent conditions, we showed that 
point (a) is always stable for $\Omega_{\phi} < 1$. 
Hence the solutions do not exit from the scaling regime 
to the epoch of cosmic acceleration driven by point (b).
This situation is analogous to what happens for Quintessence 
with the potential $V(\phi)=V_0e^{-\beta \phi}$.

To realize scaling radiation and matter eras followed by 
the epoch of cosmic acceleration, we considered 
two exponential potentials 
$V(\phi)=V_1e^{-\beta_1 \phi}+V_2e^{-\beta_2 \phi}$ 
with $\beta_1 \gg {\cal O}(1)$ and $\beta_2 \lesssim 
{\cal O}(1)$ in Sec.~\ref{Sec:Cosmology}. 
In this case, the first potential 
$V_1e^{-\beta_1 \phi}$ gives rise to 
the scaling radiation and matter critical points 
(a1) and (a2) given by Eqs.~(\ref{a1}) and (\ref{a2}), 
respectively. The density parameter $\Omega_{G_3}$ 
arising from the cubic coupling can provide an important 
contribution to the total field density parameter 
$\Omega_{\phi}$ in the scaling regime. 
Finally, the solutions enter the epoch of cosmic acceleration 
by approaching the critical point (b) arising from the second 
exponential potential $V_2e^{-\beta_2 \phi}$. 
 
For the model proposed in Sec.~\ref{Sec:Cosmology}, 
we have found some interesting features which make the model 
appealing as a viable dark energy candidate.  
Our findings are summarized below.
\begin{enumerate}
\item \textit{Scaling solutions followed by the dark energy attractor}: 
As illustrated in Fig.~\ref{M1densityics}, we have numerically confirmed that
the solutions first enter 
the scaling radiation regime and finally approach 
the dark energy attractor.
The duration of the scaling matter era ($w_\phi \simeq 0$) depends on how similar the values of
$\Omega_{\phi}$ on the critical points (a1) and (a2) are
 (see Fig.~\ref{wDE}).
Even if $\Omega_{G_3}$ dominates over the standard
field density parameter $\Omega_{G_2}$ in scaling 
radiation/matter dominated epochs, the former tends to be suppressed 
at low redshifts (see Fig.~\ref{DensityM2}).
\item \textit{Parameter space:} 
We derived theoretically consistent conditions for points (a1), (a2), (b) and showed the existence of viable parameter space 
in Fig.~\ref{stability} for some values of $\beta_1, \beta_2$.
We also considered four different G3 models existing inside 
the viable parameter space and showed that the quantities 
$Q_s$ and $c_s^2$ remain positive throughout the cosmological evolution (see Fig.~\ref{csQs}). 
The G3 models can be also cosmologically viable even 
for $\beta_2^2>2$ (such as the model M4 shown in Fig.~\ref{DensityM4}), 
while this is not the case 
for Quintessence. Hence the cubic coupling offers
the possibility for realizing the cosmic acceleration 
even for the steep second exponential potential 
$V_2e^{-\beta_2 \phi}$ with $\beta_2^2>2$.
\item \textit{Dark energy equation of state 
today, $w_{\phi}^{(0)}$}: 
For all the G3 models presented in Table \ref{Models}, we found that $w_{\phi}^{(0)}$ is closer to $-1$ in comparison 
to Quintessence. Then, these G3 models are in better agreement 
with recent observational data.
Since the evolution of $w_{\phi}$ after the onset of the 
matter-dominated epoch is also different among 
different G3 models, 
it will be also possible to observationally distinguish between them with future high-precision data.
\item \textit{Early-time dark energy density}: 
The BBN and CMB bounds (\ref{earlyDEconst}) and 
(\ref{earlyDEconst2}) on the density parameter 
$\Omega_{\phi}$ put further constraints on the model parameters, but a wide range of viable 
parameter space is still left (see Fig.~\ref{stability}).
There exist G3 models like M3 given in Table \ref{Models} 
where $\Omega_{\phi}$ around the redshift $z=50$ 
is smaller than that in Quintessence by one order of magnitude.
\item \textit{Linear perturbations:} We gave a hint to the expected modification of gravity on scales relevant to 
the growth of large-scale structures. 
Under the quasi-static approximation for modes deep 
inside the sound horizon, we showed that the parameters 
$\mu$ and $\Sigma$, which are related to the Newtonian 
and weak lensing gravitational potentials respectively, 
are given by Eq.~(\ref{mucon}). 
As we observe in Fig.~\ref{fig:mu}, the deviation of $\mu$ from 1 induced by the cubic coupling $G_3$ tends to increase
for lower redshifts. 
Thus, the G3 models give rise to observational signatures 
that can be distinguished from Quintessence.
\end{enumerate}

We conclude that the proposed model reveals very interesting features for realizing the late-time acceleration and alleviating the coincidence problem. It will be of interest to further analyze its phenomenology and  to put observational constraints on the model parameters. In particular, although the density associated 
with cubic interactions is typically suppressed at low redshifts 
in our model, the galaxy-ISW correlation data may put 
further bounds on the allowed parameter space.
The investigation of other forms of function $G_3(Y)$ could also exhibit interesting phenomenology. 
It would be also relevant to study the cosmology in the presence of couplings between 
the scalar field and matter (which are present for the original 
construction of scaling solutions in K-essence \cite{Piazza:2004df,Tsujikawa:2004dp}).
These issues are left for future works.

\begin{acknowledgements}
The research of NF and NJN is supported by Funda\c{c}\~{a}o para a  Ci\^{e}ncia e a Tecnologia (FCT) through national funds  (UID/FIS/04434/2013) and by FEDER through COMPETE2020  (POCI-01-0145-FEDER-007672). NJN  is also supported by an FCT Research contract, with reference IF/00852/2015. 
ST is supported by the Grant-in-Aid 
for Scientific Research Fund of the JSPS No.~16K05359 and 
MEXT KAKENHI Grant-in-Aid for 
Scientific Research on Innovative Areas ``Cosmic Acceleration'' (No.\,15H05890). 
\end{acknowledgements}

\appendix
\section{Eigenvalues of the critical points}
\label{Eigenvalues}

The stabilities of critical points ($x_c, y_c$) presented in Table \ref{critp} 
are known by considering homogenous perturbations 
($\delta x, \delta y$) around them, such that 
\be
x = x_c + \delta x\,,\qquad 
y = y_c +  \delta y\,.
\ee
Substituting these expressions 
into Eqs.~(\ref{xprime}) and (\ref{yprime}), 
the perturbations, at linear order, obey the differential equations
\be
\frac{d}{dN} 
\begin{pmatrix} \delta x \\ \delta y \end{pmatrix} 
= \mathcal{M}
\begin{pmatrix} \delta x \\ \delta y \end{pmatrix}\,,
\label{pereq}
\ee
where the Jacobian matrix $\mathcal{M}$ is given by
\be
\mathcal{M} = 
\begin{pmatrix}
\frac{\partial x'}{\partial x} & \frac{\partial x'}{\partial y} \\
\frac{\partial y'}{\partial x} & \frac{\partial y'}{\partial y}
\end{pmatrix}_{(x=x_c , y=y_c)}\,.
\ee
The general solutions to $\delta x$ and $\delta y$ can be expressed 
as the linear combinations of two terms $e^{\mu_1 N}$ and 
$e^{\mu_2 N}$, where $\mu_1$ and $\mu_2$ are the eigenvalues 
of $\mathcal{M}$.  
As we explained in Sec.~\ref{Sec:DynSys}, the stabilities of fixed points
are determined by the signs of $\mu_1$ and $\mu_2$.
For the critical points presented in Table \ref{critp}, we obtain 
the following eigenvalues:
\begin{itemize}
\item Point (a):	
\be	
\mu_{1 \atop 2} = \frac{3}{4} (\gamma-2)
\left[ 1 \pm \sqrt{1-\frac{8(1-\Omega_{\phi})
[\gamma + 2 A (\beta-\gamma\lambda)]}
{(2-\gamma)[1+2A(3A-\lambda)]}}\right]\,,
\label{aeigen}
\ee
where $\Omega_{\phi}$ is given by Eq.~(\ref{Omea}).
\item Point (b):		
\ba
\mu_1 &=& -3\gamma + \frac{\beta(\beta-6A)}{1+A(\beta-2\lambda)}, 
\label{beigen1}\\
\mu_2 &=& \frac{\beta^2-6 +12A(\lambda-\beta)}{2[1+A(\beta-2\lambda)]}\,.
\label{beigen2}
\ea
\item Point (c):
\ba
\mu_1 &=& -\frac{3}{2} \left(2-\gamma \right)\,,
\label{ceigen1}\\
\mu_2 &=& \frac{3\gamma}{2} + \frac{3A\beta}{1-2A\lambda}\,.
\label{ceigen2}
\ea
\item Points (d1) and (d2):
\ba
\mu_1 &=& 3(2-\gamma), \\
\mu_2 &=&  \frac{6 + 6 A(\beta-2 \lambda) \mp \sqrt{6} \beta  \sqrt{1+2A(3A-\lambda)}}{2(1-2A \lambda)}\,,
\nn\\
&&
\ea
where the $(-)$ and $(+)$ signs of $\mu_2$ correspond to 
the critical points (d1) and (d2), respectively.
\end{itemize}


\end{document}